\documentclass[11pt,a4paper]{article}
\usepackage{diagbox}
\usepackage{svg}
\usepackage{jheppub}
\usepackage{braket}
\usepackage{amsmath}
\usepackage{bbold}
\usepackage{cleveref}
\usepackage{amsfonts, amsthm}
\usepackage{mathrsfs}
\usepackage[normalem]{ulem}
\usepackage{verbatim}
\usepackage{amssymb}
\usepackage{amsthm}
\usepackage{inputenc, array}
\usepackage{textcomp}
\usepackage{upgreek}
\usepackage{appendix}
\usepackage[dvipsnames]{xcolor}
\usepackage{epsfig}
\usepackage{graphicx}
\usepackage{array}
\usepackage{multirow}
\usepackage{makecell}
\usepackage{pifont}
\newcolumntype{C}[1]{>{\centering\arraybackslash}p{#1}}
\newcommand{\n}{\nu}
\newcommand{\m}{\mu}
\newcommand{\nn}{\nonumber}

\newcommand{\xcentcolon}

\usepackage{color}
\newcommand{\be}[1]{\begin{equation}\label{#1} }
\newcommand{\ee}{\end{equation}}
\newcommand{\bea}[1]{\begin{eqnarray}\label{#1} }
\newcommand{\eea}{\end{eqnarray}}

\renewcommand{\O}{{\mathcal{O}}}
\renewcommand{\L}{{\mathcal{L}}}

\newcommand{\lb}{\left(}
\newcommand{\rb}{\right)}

\renewcommand{\a}{\alpha}

\renewcommand{\b}{\beta}
\renewcommand{\t}{\tau}
\newcommand{\s}{\sigma}

\theoremstyle{plain} % For theorems, lemmas, corollaries (italic by default)

\theoremstyle{definition} % For definitions (non-italic)

\definecolor{myblue}{RGB}{11,83,148}
\newcommand{\xf}{x^\phi}
\newcommand{\xff}{x^\varphi}
\newcommand{\xt}{x^t}
\newcommand{\xtt}{x^T}
\newcommand{\xr}{x^\rho}
\newcommand{\yf}{y^\phi}

\newcommand{\xd}{\dot{x}}

\title{Strings near BTZ black holes: A Carrollian Chronicle}

\author[a]{Aritra Banerjee,}
\emailAdd{aritra.banerjee@pilani.bits-pilani.ac.in}
\author[b]{Arkachur Bhattacharya,}
\emailAdd{arkachurb25@iitk.ac.in}
\author[b]{Sharang Rajesh Iyer,}
\emailAdd{siyer23@iitk.ac.in}
\author[a,b,c]{Ansh Mishra,}
\emailAdd{anshmishra471@gmail.com}
\author[b,d]{Priyadarshini Pandit}
\emailAdd{priyadarshini.pandit@tifr.res.in}
\affiliation[a]{Birla Institute of Technology and Science, Pilani Campus, Pilani, Rajasthan 333031, India.}
\affiliation[b]{Indian Institute of Technology Kanpur, Kanpur 208016, India.}
\affiliation[c]{Indian Institute of Science Education and Research, Mohali, Punjab 140306, India.}
\affiliation[d]{Tata Institute of Fundamental Research, Homi Bhabha Rd, Mumbai 400005, India.}
\preprint{}
\abstract{The BTZ black hole provides a tractable (2+1)-dimensional example for investigating string dynamics in curved spacetime. However, a systematic and robust analysis of the solution space of strings in the near-horizon region of BTZ black holes remains elusive in the literature. This work aims to fill this gap by employing the \textit{string–Carroll expansion}. This formalism provides a natural setting for working with the near-horizon region, because near-horizon expansions for non-extremal black holes match string-Carroll expansions. Using this formalism, and expanding the string action and pullback fields in powers of an effective speed of light, we study the dynamics of closed bosonic strings in the near-horizon, non-extremal BTZ spacetime. Our approach classifies the general characteristics and further reveals some novel features of the families of string solutions.}

\begin{document}
\maketitle
	\flushbottom
\newpage

\section{Introduction}

Three-dimensional Einstein gravity serves as a simple yet valuable toy model for studying quantum gravity \cite{DESER1984220,Deser1988,tHooft1988,WITTEN198846,Witten:1989sx} with properties extendable to higher dimensions. Although three-dimensional gravity is characterised by the absence of propagating degrees of freedom \cite{Carlip:1995qv,Compere:2018aar} and lacks a Newtonian limit \cite{JDBarrow_1986}, it remains a remarkably powerful framework, especially in the context of the anti-de Sitter/conformal field theory (AdS/CFT) correspondence \cite{Maldacena:1997re,Brown:1986nw}. A central example in this setting is provided by the Ba\~nados-Teitelboim-Zanelli (BTZ) black holes \cite{Banados:1992wn,PhysRevD.48.1506,Carlip:1995qv,Carlip:2005zn}. The BTZ black hole is an exact solution to Einstein's field equations in 2+1 dimensions with a negative cosmological constant. Since the curvature in 2+1 dimensions is constant, the BTZ black hole lacks a curvature singularity at the origin. Nevertheless, these black hole solutions share many properties with their higher-dimensional counterparts, but are analytically more tractable and hence serve as valuable guides for studying the generic properties of black hole spacetimes.
\medskip

Geometrically, an electrically neutral BTZ black hole can be constructed as an orbifold of the universal covering space of $\text{AdS}_3$ \cite{Banados:1992wn,PhysRevD.48.1506, Carlip:2005zn, Kraus2008}. In other words, it is obtained by identifying points in AdS$_3$ under a discrete subgroup of its isometry group, leading to a quotient geometry that retains the local AdS$_3$ structure but has distinct global properties, including the presence of a horizon(s). The metric of the BTZ black hole is asymptotically AdS$_3$ which can also be described as an $SL(2,\mathbb{R})$ group manifold \cite{Maldacena:2000hw}. Thus, a string propagating in an AdS$_3$ target spacetime is simply described by a Wess-Zumino-Witten (WZW) action \cite{Gepner:1986wi} on a $SL(2,\mathbb{R})$ group manifold. The seminal set of works \cite{Maldacena:2000hw,Maldacena:2000kv,Maldacena:2001km} by Maldacena and Ooguri provides us with the details of the connection between strings in an AdS$_3$ target spacetime and the $SL(2,\mathbb{R})$ WZW model. It has been shown that the $SL(2,\mathbb{R})$ WZW action for string theory is a Polyakov action with a BTZ black hole target space metric along with a Kalb-Ramond background field \cite{Hemming:2001we}. The affine current algebra of the WZW model, together with the Virasoro constraints, controls the full stringy spectrum on AdS$_3$, and features such as spectral flow, winding and the long-string sector play an essential role in microscopic accounts of the BTZ entropy \cite{Giveon:1998ns, Kutasov:1999xu}.
\medskip

Motivated by three decades of progress in related directions, in the current work, we study closed strings in a BTZ black hole background, focusing on a special limit where the string is very close to the horizon. Rather than employing standard worldsheet methods, we use a novel limiting procedure, which we now elaborate on.  While earlier works \cite{Bars:1994sv,PhysRevD.13.2364,deVega:1987um,Lowe:1994ah} have explored related questions, our approach relies on a different framework to capture near-horizon string dynamics. The absence of a non-degenerate metric on the whole of the near-horizon spacetime necessitates the use of non-Lorentzian structures. In particular, here we need to invoke the so-called \textit{string-Carroll geometry}.
\medskip

Carrollian physics, as well known now, arises as the vanishing speed of light limit of relativistic theories \cite{Leblond65,SenGupta:1966qer,Henneaux:1979vn}. It has recently drawn significant attention owing to its appearance across diverse physical settings \footnote{See \cite{Bagchi:2025vri} for a recent review on multiple facets of Carrollian physics and a mostly exhaustive list of references.}. Carrollian structures emerge on the event horizons of generic black holes \cite{Henneaux:1979vn,Donnay:2019jiz} (see also \cite{Fontanella:2022gyt, blair2025carrollgeometrymeetssitter} for related discussions on non-Lorentzian structures of near-horizon geometries) and also play a central role in certain condensed matter systems such as systems with flat-bands \cite{Bagchi:2022eui,Ara:2024fbr}, phase separation regions \cite{Biswas:2025dte} and fractons \cite{Bidussi:2021nmp, Figueroa-OFarrill:2023vbj}. These structures appear in the hydrodynamics of the quark–gluon plasma \cite{Bagchi:2023ysc,Bagchi:2023rwd,Kolekar:2024cfg}, and even feature in cosmological scenarios \cite{deBoer:2021jej}. Conformal extensions of Carrollian symmetries have further been proposed as holographic duals of asymptotically flat spacetimes (AFS) \cite{Bagchi:2010zz,Barnich:2012aw,Bagchi:2012yk,Bagchi:2012xr,Barnich:2012xq,Bagchi:2012cy,Bagchi:2016bcd,Donnay:2022aba,Bagchi:2022emh,Donnay:2022wvx,Bagchi:2023fbj}. Interestingly, the same symmetries also emerge on the worldsheet of tensionless (null) strings as residual gauge symmetries \cite{Bagchi:2013bga,Bagchi:2015nca}.  
\medskip 

The Carroll expansion of a theory is one where a relativistic theory is expanded in orders of small but finite $c$, instead of setting $c$ to zero \cite{deBoer:2021jej,deBoer:2023fnj,Bagchi:2023cfp,Bagchi:2024rje, Hansen:2021fxi}. In usual relativistic theories, $c$ appears along with the time coordinate. Hence, in the Carroll expansion of field theories, the temporal direction is singled out, which, after the expansion, turns out to be a null direction. This is referred to as the \textit{particle-Carroll expansion}. We now consider a generalisation of the Carroll expansion where there are two, instead of a single null direction, the \textit{string Carroll expansion}. As we will review in a later section, the near-horizon expansion of non-extremal black holes can be recast as a string Carroll expansion. This was first noticed in the context of Schwarzschild black holes in four-dimensional asymptotically flat spacetimes (AFS) in \cite{Bagchi:2023cfp, Bagchi:2024rje}, where string solutions were studied from the point of view of an observer at infinity. In the current paper, we extend the analysis to three-dimensional BTZ black holes. Apart from the obvious distinctions between $3d$ and $4d$ horizons, there arise differences due to BTZ being asymptotically AdS vs Schwarzschild being AFS. We also extend our analysis to the rotating BTZ case and see further intricacies.
\medskip

One could immediately ask, why would we want to use such a formalism? The reason is that in the near-horizon limit of a non-extremal black hole, the geometry factorises locally into the product of a longitudinal two-dimensional Rindler spacetime (spanned by the near-horizon time and radial directions) and a transverse space at fixed horizon radius \cite{Townsend:1997ku}. In this limit, the whole of the near-horizon spacetime degenerates, and hence attempting to understand usual relativistic quantum fields as well as strings in this degenerate background becomes impossible in the conventional Lorentzian sense. As we will elaborate below, the string-Carroll formalism naturally takes into account the degenerate nature of the near-horizon expansion, and we systematically get Carrollian strings. Motivated by these subtleties, we build on a robust framework for studying the near-horizon geometry of non-extremal black holes in terms of string-Carroll expansions. 
\medskip

In field theoretic investigations \cite{Henneaux:2021yzg, deBoer:2021jej, deBoer:2023fnj}, the particle-Carroll expansion of the relativistic field theories leads to a leading-order (LO) theory, known as the electric Carroll theory and a next-to-leading order (NLO) theory, also known as the magnetic Carroll theory.
As mentioned, \cite{Bagchi:2023cfp,Bagchi:2024rje} have explored similar electric and magnetic sectors in the context of string theory near a Schwarzschild black hole. In a slightly different vein as in field theory, string-Carroll expansion of the Polyakov action in our case in particular leads to a LO theory which describes magnetic strings. In this theory, the target spacetime is degenerate while the worldsheet theory is Lorentzian. We will also show that magnetic strings can be obtained from the string-Carroll expansion of the relativistic phase-space theory as well. In contrast, we obtain the electric strings only from the string-Carroll expansion of the relativistic phase-space theory. The LO electric theory in this case corresponds to a theory of null strings appearing with a tensile deformation term. We will elaborate on these sectors of string theory near a BTZ black hole in the forthcoming sections.
\medskip

One should note here, the BTZ black hole differs from the Schwarzschild black hole in topological aspects. The LO near-horizon geometry of a Schwarzschild black hole is a two-sphere $S^2$, while the NLO consists of a 2D Rindler spacetime together with a metric on the angular directions. However, the near-horizon geometry of a BTZ black hole is a sphere $S^1$ at the LO and a combination of a 2D Rindler spacetime and a metric on the angular direction at the NLO. As we will see in this work, strings near a black hole see an enlarged sphere as compared to the 2D Rindler in both cases. Therefore, the topological differences between $S^2$ and $S^1$ become important for classical solutions. In essence, the solution space of the Virasoro constraints and the equations of motion of the string near a BTZ black hole becomes more constrained, but at the same time more analytically tractable!
\medskip

This paper is structured as follows. In section \ref{Near-horizon expansion of BTZ black hole is string Carroll}, we show that the near-horizon geometry of a static and rotating non-extremal BTZ black hole is string-Carroll. We also provide a self-contained review on string-Carroll geometry and the string-Carroll expansion of a Lorentzian geometry. We further review BTZ black holes and go on to map the near-horizon geometries of static and rotating non-extremal BTZ black holes to the string-Carroll geometry. Section \ref{BTZ black hole: Magnetic String} deals with the properties of magnetic strings near BTZ black holes via two pathways. Firstly, by expanding the Polyakov action and secondly using the phase space string action. In section \ref{BTZ black hole: Electric Strings}, we discuss electric version of these strings, starting by first showing that the residual symmetries of the worldsheet close the $\text{BMS}_3$ algebra, despite having a finite tension piece \footnote{While this manuscript was in preparation, an independent work \cite{Figueroa-OFarrill:2025njv} appeared on the arXiv reporting the same result.}. Then we discuss the classification of electric strings near BTZ black holes. The generic structures in the solution space of strings for BTZ black holes are discussed in section \ref{Features of three and four dimensions}. Section \ref{Concluding Remarks and Future Directions} provides a summary of our findings along with prospective avenues for future exploration. Finally, the appendices contain some more interesting discussions beyond the scope of the main text. In Appendix \ref{appendix:null string} we discuss the solution space of null strings, that is, when the tension is set to zero, while in Appendix \ref{app:Particle Carroll expansion of relativistic strings} we elaborate on the particle Carroll expansion of string theory.

\section{Near-horizon BTZ black hole: String Carroll expansions}\label{Near-horizon expansion of BTZ black hole is string Carroll}
In this section, we first review the string Carroll expansion of \cite{Bagchi:2023cfp,Bagchi:2024rje} which stemmed from the approach taken for string Newton-Cartan expansion in \cite{PhysRevLett.128.021602, Hartong:2022dsx} and then analyse the near-horizon expansion of the BTZ black holes for both static and rotating (non-extremal) cases. We show that for both the static and rotating cases, the near-horizon geometry admits a string Carroll expansion, as introduced in \cite{Bagchi:2023cfp}.

\subsection{Preliminaries: String Carroll}\label{String Carroll expansion}

In a general setting, instead of setting the speed of light $c$ to zero from the outset, it is advantageous to consider a perturbative expansion in powers of $c$, treating it as small but finite. The expanded string Carroll metric $g$, for a $(d+2)$-dimensional Lorentzian spacetime, takes the following general form 
\begin{subequations}\label{eq:string-Carroll metric}
\begin{align}\label{scexp}
    g_{\mu\nu}&=\Omega_{\mu\nu}+c^2\tau_{\mu\nu}+c^2\Theta_{\mu\nu}+\mathcal{O}(c^4),\\
g^{\mu\nu}&=c^{-2}\upsilon^{\mu\nu}+\tilde{\Omega}^{\mu\nu}+\mathcal{O}(c^2).
\end{align}
\end{subequations}
Here $\Omega_{\mu\nu}$ carries the signature $(0,0,+,+,\cdots)$ whereas the signature of $\tau_{\mu\nu}$ and $\upsilon^{\mu\nu}$ is $(-,+,0,0,\cdots)$. The exact form of $\tilde{\Omega}^{\mu\nu}$ will be made clear in the vielbeine formalism below. Unlike the particle Carroll expansion, which singles out the temporal direction, the string-Carroll expansion inherently involves scaling of two distinguished directions. We will refer to these two directions as the longitudinal directions, whereas the others will be called transverse directions. Following this framework, the flat version of the string-Carroll metric is given as
\begin{equation}
   g(c) = -c^2\, (dx^0)^2 + c^2\, (dx^1)^2  + dx^idx^i.
\end{equation}	
In the vielbeine formalism, the string-Carroll geometry is split in terms of longitudinal and transverse vielbeine\footnote{Also known as pre-Carrollian variables.} as follows
\begin{align}
  g_{\mu\nu} = c^2 \eta_{AB} \mathcal{T}^A_\mu \mathcal{T}^B_\nu + \delta_{ij}\mathcal{E}^i_\mu\mathcal{E}^j_\nu,\quad g^{\mu\nu} = \frac{1}{c^2} \eta^{AB} \mathcal{V}^\mu_A \mathcal{V}^\nu_B + \delta^{ij}\mathcal{E}_i^\mu\mathcal{E}_j^\nu,
\end{align}	
where Greek indices run over the entire spacetime i.e, $\mu,\nu=\{0,1,\dots,d+1\}$,  $A, B=\{0,1\}$ are the tangent space indices in the longitudinal directions and $i,j=\{2,\ldots,d+1\}$ are the tangent space indices in the transverse directions. Here $\eta_{AB}=\text{diag}(-1,1)$ is the longitudinal tangent space metric and $\delta_{ij}=\text{diag}(1,\ldots,1)$ constitutes the metric for the transverse tangent space. The invertability condition of the full spacetime metric, $g_{\mu\rho} g^{\rho\nu} = \delta^\nu_\mu$, imposes the following conditions on the vielbeine
 \begin{align}
\mathcal{T}^A_\mu \mathcal{V}^\mu_B = \delta^A_B,~~~\mathcal{E}_\mu^i\mathcal{E}^\mu_j=\delta^i_j,~~~\mathcal{T}^A_\mu \mathcal{E}^\mu_i = \mathcal{V}^\mu_A\mathcal{E}_\mu^i = 0,~~~\mathcal{T}^A_\mu \mathcal{V}^\nu_A+\mathcal{E}_\mu^i\mathcal{E}^\nu_i=\delta^\nu_\mu.
\end{align}	
The expanded string-Carroll structure \eqref{eq:string-Carroll metric} is obtained by expanding the vielbeine in powers of $c^2$ as follows
\begin{subequations}
\begin{align}
\mathcal{T}^A_\mu &= \tau^A_\mu +c^2m_\mu^A +\mathcal{O}(c^4),~~~
\mathcal{E}_\mu^i=e_\mu^i+c^2\varpi^i_\mu+\mathcal{O}(c^4),\\
\mathcal{V}_A^\mu &=\upsilon_A^\mu+c^2M^\mu_A+\mathcal{O}(c^4),~~~
\mathcal{E}^\mu_i=e^\mu_i+c^2\pi_i^\mu+\mathcal{O}(c^4).
\end{align}    
\end{subequations}
Having expanded the vielbeine, the expanded string-Carroll metric takes the form \eqref{eq:string-Carroll metric}, where
\begin{subequations}
\begin{align}
\Omega_{\mu\nu}&=\delta_{ij}e^i_\mu e^j_\nu,~~~\tau_{\mu\nu}=\eta_{AB}\tau_\m^A\tau_\n^B,~~~\Theta_{\mu\nu}=2\delta_{ij}e^i_{(\mu}\varpi^j_{\n)}\\
\upsilon^{\mu\nu}&=\eta^{AB}\upsilon^\m_A\upsilon^\n_B,~~~\tilde{\Omega}^{\mu\nu}=\delta^{ij}e^\mu_ie^\nu_j+2\eta^{AB}\upsilon^{(\mu}_AM^{\nu)}_B.
\end{align}
\end{subequations}

The causal structure of string Carroll geometry and the transformations of the vielbeine under local tangent space transformations have been discussed at length in \cite{Bagchi:2024rje}, which we refer the interested readers to. In what follows, we perform the near-horizon expansion of the BTZ black hole and establish a correspondence between the string-Carroll geometry and the near-horizon geometry of the BTZ black hole.

\subsection{BTZ black hole: A quick review}\label{BTZ black hole: A quick review}
The BTZ black hole is a non-trivial solution to Einstein’s field equations in three-dimensional spacetime with a negative cosmological constant ($\Lambda = -1/\ell^2$). It describes an asymptotically $AdS_3$ black hole with mass and angular momentum \cite{Banados:1992wn,PhysRevD.48.1506}.
The BTZ black hole metric in the Boyer-Lindquist-like coordinate system is given as\footnote{In units where $c = 1$ and the gravitational constant, $G = 1/8$.} \cite{Compere:2018aar}
	
	\begin{equation}\label{general btz}
	ds^2 = -N^2(r)\, dt^2 + \frac{dr^2}{N^2(r)} + r^2 \left(N^\phi(r) dt + d\phi \right)^2,
	\end{equation}
	where $N^2(r)$ and $N^\phi(r)$ are the lapse and the angular drag functions, respectively, and are given as follows
	\begin{equation}
    N^2(r) = -M + \frac{r^2}{\ell^2} + \frac{J^2}{4r^2}\,,\quad N^\phi (r)= -\frac{J}{2r^2}\,.
	\end{equation}
The integration constants $M$ and $J$ are Noether-Wald conserved charges and are interpreted as the total mass and total angular momentum of the black hole. The zeroes of the lapse function define the event horizons of the BTZ black hole. The outer and the inner horizons are located at radii $r_\pm$. In terms of parameters $M$ and $J$, the radii are
\begin{equation}\label{event horizon radii}
	r_\pm = \ell \left[ \frac{M}{2} \left( 1 \pm \sqrt{ 1 - \left( \frac{J}{M\ell} \right)^2 }  \right)  \right]^\frac{1}{2}.
\end{equation}
For real and positive radii, $M\in \mathbb{R}^+$ and $|J|\leq M\ell$. The second condition gives rise to two types of BTZ black holes, namely, static and rotating. A static BTZ black hole has $J=0$, whereas a rotating BTZ black hole has $0<|J|\leq M\ell$. A special case of the rotating BTZ black hole is the extremal BTZ black hole, which has $|J|=M\ell$. For the extremal scenario, the event horizons of the black hole merge to form a single horizon. 
\medskip

To analyse the near-horizon expansions of BTZ black hole geometries, we restrict our attention to the static and rotating non-extremal cases. The horizons\footnote{Outer horizon in the case of non-extremal BTZ black holes.}, in both cases, can be probed by introducing a small dimensionless parameter $\epsilon$ through the following coordinate transformation
\begin{equation}\label{nhexp}
        r = r_h + \epsilon \rho.
\end{equation}
 We will show that, up to $\mathcal{O}(\epsilon)$, the string-Carroll expansion as done in \eqref{eq:string-Carroll metric} and near-horizon expansions of the black holes can be mapped, provided the effective distance parameter $\epsilon$ is identified with $c^2$.

\subsection{Near-horizon expansion of static BTZ black hole}\label{Near-horizon expansion of static BTZ black hole}
The metric for the static BTZ black hole is given as
\begin{equation}\label{static btz}
	ds^2 = -\left( -M + \frac{r^2}{\ell^2} \right) dt^2 + \left( -M + \frac{r^2}{\ell^2} \right)^{-1} dr^2 + r^2 d\phi^2.
\end{equation}
The event horizon of the static BTZ black hole is located at $r_h = \ell \sqrt{M}$. Applying the coordinate transformation \eqref{nhexp} to the static BTZ metric \eqref{static btz} yields a near-horizon expansion of the geometry as follows
\begin{equation}\label{static expansion 1}
	ds^2 = r_h^2 \, d\phi^2 + \epsilon \left( -\frac{\rho}{\alpha} dt^2 + \frac{\alpha}{\rho} d\rho^2 + 2r_h \rho \, d\phi^2 \right) + \O(\epsilon^2)\,,
\end{equation}
where $\alpha = \frac{\ell^2}{2r_h}$. In the above expansion \eqref{static expansion 1}, $t$ and $\rho$ span the longitudinal directions, whereas the transverse direction is spanned by $\phi$. Under the coordinate transformation $\rho=\frac{\eta^2}{4\alpha}$, the longitudinal sector of the metric \eqref{static expansion 1}, evaluated at order $\O(\epsilon)$, takes the usual form of a two-dimensional Rindler metric
\begin{equation}
    ds^2 = -\left(\frac{\eta}{2\alpha}\right)^2 dt^2 + d\eta^2 \,.
\end{equation}

Comparing \eqref{static expansion 1} with the string-Carroll metric expansion \eqref{eq:string-Carroll metric}, and identifying $\epsilon$ with $c^2$, we obtain the following map between the two expansions
    \begin{subequations}\label{static expansion components}
        \begin{align}
		\Omega_{\mu\nu} dx^\mu dx^\nu =&\; r_h^2 d\phi^2,\\
		\eta_{AB} \tau_\mu^A \tau_\nu^B dx^\mu dx^\nu =&\; -\frac{\rho}{\alpha} dt^2 + \frac{\alpha}{\rho} d\rho^2, \\
		\Theta_{\mu\nu} dx^\mu dx^\nu =&\; 2 r_h \rho d\phi^2 \,.
	\end{align}
    \end{subequations}    
In the next subsection, we establish a similar map between the rotating non-extremal BTZ black hole and the string-Carroll expansion.

\subsection{Near-horizon expansion of rotating (non-extremal) BTZ black hole}\label{Near-horizon expansion of rotating (non-extremal) BTZ black hole}

The map between the string-Carroll and the near-horizon expansions for the rotating case is best established by going to the ``co-rotating" frame, where the angular velocity at the horizon is effectively absorbed, simplifying the near-horizon geometry. The required coordinate transformations are
\begin{equation}\label{non extremal coordinate transform}
t\to T=\left(1+\frac{J^2\alpha_+}{2r_+^3}  \right)t - \left(\frac{J \alpha_+}{r_+}\right) \phi,~~~\phi \to \varphi= \phi - \frac{J}{2r_+^2} t,
\end{equation}
where $r_\pm$ are as given in \eqref{event horizon radii} and we have defined $\alpha_+$ as
\begin{equation} \label{alpha plus}
	    \alpha_+ = \lb \frac{2r_+}{\ell^2} - \frac{J^2}{2r_+^3} \rb^{-1} = \frac{\ell^2 r_+}{2 \left(r_+^2-r_-^2\right)} \, .
        %\qquad \alpha_+ \xrightarrow{\hspace{1mm} J \to 0 \hspace{1mm}} \alpha.
\end{equation}

\begin{comment}
    Note that $\varphi$ is still an angular coordinate, since 
	\[ \phi \sim \phi + 2n\pi, \; n \in \mathbb{Z} \quad
	\implies 
	\varphi \sim \varphi + 2n\pi. \]
\end{comment}

In these new coordinates, the metric of a non-extremal rotating BTZ black hole becomes diagonal. The form of the metric in terms of the two horizon radii is given below
\begin{equation}
    g = -\frac{ \left(r^2-r_+^2\right) \left(r_+^2-r_-^2\right)}{\ell^2 r_+^2}dT^2 + \frac{ \ell^2 r^2}{\left(r^2-r_-^2\right) \left(r^2-r_+^2\right)} dr^2 + \frac{ r_+^2 \left(r^2-r_-^2\right) }{r_+^2-r_-^2} d\varphi^2 \,.
\end{equation}

The reader may refer to \cite{Krishnan:2009kj} to know the subtleties about diagonalised coordinate systems in rotating BTZ black holes. We can focus on the near-horizon region of the rotating BTZ black hole by using the same coordinate transformation \eqref{nhexp}, now about $r_+$. The near-horizon metric takes the familiar form
\begin{equation}\label{non extremal expansion 1}
	g = r_+^2 d\varphi^2 + \epsilon 
		\left[
		-\left(\frac{\rho}{\alpha_+}\right) dT^2 
		+ \left(\frac{\alpha_+}{\rho}\right) d\rho^2 
		+ \left( 2r_+ + \frac{\alpha_+ J^2}{r_+^2} \right) \rho d\varphi^2
		\right] 
		+ \mathcal{O}(\epsilon^2)\,.
\end{equation}
Finally, we map \eqref{non extremal expansion 1} to the string-Carroll expansion \eqref{eq:string-Carroll metric} in 2+1 dimensions as
\begin{subequations}\label{non extremal expansion components}
\begin{align}
		\eta_{AB} \tau_\mu^A \tau_\nu^B dx^\mu dx^\nu &= -\left(\frac{\rho}{\alpha_+}\right) dT^2 + \left(\frac{\alpha_+}{\rho}\right) d\rho^2,\\
		\Theta_{\mu\nu} dx^\mu dx^\nu &= \left( 2r_+ + \frac{\alpha_+ J^2}{r_+^2} \right) \rho\, d\varphi^2,\\
		\Omega_{\mu\nu} dx^\mu dx^\nu &= r_+^2\, d\varphi^2\,.
\end{align}
\end{subequations}
Notice that the static \eqref{static expansion 1} and the rotating \eqref{non extremal expansion 1} BTZ black holes share some similarities in their near-horizon expansions. In the case of rotating BTZ black holes, $T$ and $\rho$ are the coordinates for the longitudinal directions, whereas $\varphi$ is the coordinate for the transverse direction. At LO, i.e. $\O(\epsilon^0)$, the transverse direction is a circle. The leading-order geometry in the longitudinal directions retains its Rindler structure. Setting $J=0$ in \eqref{non extremal expansion components}, we straightforwardly recover \eqref{static expansion components} after identifying $r_+$ with $r_h$. Furthermore, the definition of the transverse coordinate $\varphi$ from \eqref{non extremal coordinate transform} makes the existence of frame dragging apparent in the near-horizon region of the rotating BTZ black hole.

\section{Strings near BTZ: Magnetic Theory}\label{BTZ black hole: Magnetic String}
Once the geometric settings have been put on the table, we can now concentrate on string solutions.    
It has been previously discussed that the magnetic limit of the near-horizon string can be obtained equivalently both from the Polyakov and the phase-space formulations of the action. 
    In this section, we explore both the pathways to derive the equations of motion for the BTZ magnetic string. We begin with the Polyakov action and subsequently proceed to the phase-space action, and then compare the resulting solutions.

\subsection{Action with the string Carroll metric}\label{Polyakov action in the string Carroll metric}

    The dynamics of a classical closed string is captured by the Polyakov action \cite{Polchinski:1998rq} given by
	
	\begin{equation}\label{Polyakov}
		S_P = - \frac{T}{2}\int  \sqrt{ -\gamma }\, \gamma^{ab} \partial_a X^\mu \partial_b X^\nu  g_{\mu\nu}\,\, d^2\xi\,.
	\end{equation} 
	Here, $T$ denotes the string tension. $X^0(\xi),\dots, X^d(\xi)$ represent the embedding of the string worldsheet into the target spacetime. The parameters $\xi = (\tau, \sigma)$ are coordinates on the two-dimensional worldsheet. Greek indices $\mu, \nu \in \{0, \dots, d\}$ refer to target spacetime, while Latin indices $a, b \in \{0, 1\}$ label worldsheet coordinates.\footnote{We use $X^\mu \equiv X^\mu (\xi)$, $ g_{\mu\nu} \equiv g_{\mu\nu}\big( X^\mu (\xi) \big) $, $\partial_a = \partial/\partial\xi^a$, $\partial_\mu = \partial/\partial X^\mu$.}
	\medskip
	
	We now expand \eqref{Polyakov} in analytical powers of $\epsilon$ for a general Lorentzian metric which has a small $\epsilon$ expansion of the form \eqref{static expansion components}, \eqref{non extremal expansion components}. We can, without the loss of generality, postulate that the embedding scalars $X^\mu$, worldsheet metric $\gamma_{ab}$ and the inverse of worldsheet metric $\gamma^{ab}$ have the following analytical expansion in powers of $\epsilon$.	
	\begin{subequations}
	    \begin{align}
		X^\mu(\xi) &= x^\mu(\xi) + \epsilon y^\mu(\xi) + \mathcal{O}(\epsilon^2)\,,\label{X expansion}\\
		\label{gamma}
		\gamma_{ab}(\xi) &= \gamma_{(0)ab}(\xi) + \epsilon\gamma_{(2)ab}(\xi) + \mathcal{O}(\epsilon^2)\,, \\
		\label{gamma inv}
		\gamma^{ab}(\xi) &= \Gamma_{(0)}^{ab}(\xi) + \epsilon \Gamma_{(2)}^{ab}(\xi) + \mathcal{O}(\epsilon^2) \,.
	\end{align}
	\end{subequations}	
	$\Gamma_{(n)}$ and $\gamma_{(n)}$ are symmetric for all $n \geq 0$. A direct consequence of these choices in \eqref{gamma} and \eqref{gamma inv} are	
	\begin{subequations}
	    \begin{align}\label{worldsheet metric relation}
		\Gamma_{(0)} ^{ac}\gamma_{(0)cb} &= \delta^a_b\,, \\
		\Gamma_{(2)}^{ab}\gamma_{(0)bc} & = -\Gamma_{(0)}^{ab}\gamma_{(2)bc}\,. 
	\end{align}
	\end{subequations}	
From \eqref{worldsheet metric relation}, we observe that $\Gamma_{(0)}$ is the inverse of $\gamma_{(0)}$. Using this relation and taking the limit $\epsilon \to 0$, we conclude that $\gamma_{(0)}$ must have a Lorentzian signature. Now, let us plug the above postulates into the Polyakov action \eqref{Polyakov} and examine how each term behaves in the $\epsilon$ expansion. In this framework, the pullback of the target spacetime metric admits an expansion in $\epsilon$. The expansion and its various components are given as follows
\begin{align}\label{metric pullback expansion}
g_{ab} := &~\partial_a X^\mu \partial_b X^\nu g_{\mu\nu}(X) = \Omega_{ab}(x) + \epsilon \hat{\Theta}_{ab}(x,y) +\mathcal{O}(\epsilon^2)\,,
\end{align}
where 
\begin{subequations}
\begin{align}
\hat{\Theta}_{ab}(x,y)=&~\tau_{ab}(x)+ \Theta_{ab}(x) + 2 \Omega_{\mu\nu}(x)\partial_{(a}x^\mu\partial_{b)}y^\nu + \partial_a x^\mu\partial_b x^\nu y^\alpha \partial_\alpha \Omega_{\mu\nu}(x)\,,\\
\Omega_{ab}(x)=&~\Omega_{\mu\nu}(x)\partial_ax^\mu\partial_bx^\nu\,,~~\tau_{ab}(x)=\tau_{\mu\nu}(x)\partial_ax^\mu\partial_bx^\nu\,,~~
\Theta_{ab}(x)=\Theta_{\mu\nu}(x)\partial_ax^\mu\partial_bx^\nu\,. 
\end{align}
\end{subequations}

The Polyakov Lagrangian in \eqref{Polyakov} takes the following form of expansion	
	\begin{equation}
		\mathcal{L}_P = \mathcal{L}_{P,LO} + \epsilon \mathcal{L}_{P,NLO} + \mathcal{O}(\epsilon^2).
	\end{equation} 

    where the LO Lagrangian $\mathcal{L}_{P,LO}$ and NLO Lagrangian $\mathcal{L}_{P,NLO}$ are extracted as \footnote{While describing the worldsheet symmetries, it is argued in \cite{Bagchi:2023cfp} that $\gamma_{(0)ab}$ can be gauge fixed to $\eta_{ab}$ and $\gamma_{(2)ab}$ can be set to zero. We will follow this gauge fixing in our subsequent calculations.} 
    \begin{subequations}
    \begin{align}\label{lo polyakov}
		\mathcal{L}_{P,LO} = & -\frac{T}{2}\, \Gamma_{(0)}^{ab}(\xi)\,\Omega_{ab}\left( x\right)\sqrt{ - \gamma_{(0)} } \,, \\
		\label{nlo polyakov}
		\mathcal{L}_{P, NLO} = & -\frac{T}{2} \left[ \Gamma_{(0)}^{ab}\hat{\Theta}_{ab}(x,y)- \frac{1}{2}G_{(0)}^{abcd}\Omega_{ab}(x)\gamma_{(2)cd} \right]\sqrt{ - \gamma_{(0)} } \,.
	\end{align}
    \end{subequations}
    
Here,	$G_{(0)}^{abcd}$ is the Wheeler-DeWitt metric, defined as
	\begin{align} 
		G_{(0)}^{abcd} :=   \Gamma_{(0)}^{ac}\Gamma_{(0)}^{bd} + \Gamma_{(0)}^{ad}\Gamma_{(0)}^{bc} - \Gamma_{(0)}^{ab}\Gamma_{(0)}^{cd} \,.
	\end{align}
  Having obtained the LO \eqref{lo polyakov} and NLO Polyakov Lagrangian \eqref{nlo polyakov}, we now proceed to derive the Virasoro Constraints and the equations of motion (EOMs).

\subsection*{Virasoro Constraints}
    Variation of the LO Polyakov Lagrangian \eqref{lo polyakov} with respect to $\gamma_{(0)}$ yields the LO Virasoro constraint $T_{(0)ab}$ given as
    \begin{equation}\label{lo virasoro}
		T_{(0)ab} := \Omega_{ab}(x) - \frac{1}{2}\Gamma_{(0)}^{cd}\Omega_{cd}(x)\gamma_{(0)ab} = 0\, .
    \end{equation}
    Similarly, we obtain the NLO Virasoro constraint from the NLO Lagrangian \eqref{nlo polyakov}. Interestingly, varying $\gamma_{(2)ab}$ in the NLO Polyakov Lagrangian, yields the LO Virasoro constraint \eqref{lo virasoro}. The NLO Virasoro constraint is obtained by varying \eqref{nlo polyakov} with respect to $\gamma_{(0)ab}$, which takes the form \footnote{Terms with $\gamma_{(2)}$ are not explicitly written because we will set $\gamma_{(2)} = 0$ based on our earlier discussion.}
    \begin{equation}\label{nlo virasoro}
		T_{(2)ab} := \hat\Theta_{ab}(x,y) - \frac{1}{2}\Gamma_{(0)}^{cd}\hat\Theta_{cd}(x,y) \gamma_{(0)ab} + \left(\text{terms with } \gamma_{(2)} \right) = 0\, .
	\end{equation}
	The variations with respect to $x^\mu$ and $y^\mu$ are not explicitly derived here, as they do not offer significant insight into the general structure of the solutions. However, we will present them explicitly when we start analysing strings near a BTZ black hole from the next subsection.

\subsection{Polyakov strings near BTZ black hole: Static case}
We now apply the general formalism outlined in Section \eqref{Polyakov action in the string Carroll metric} to study string dynamics near the horizon of a static BTZ black hole to start with. Here, we derive the particular equations of motion corresponding to the LO and the NLO expansions of the Polyakov action, proceeding subsequently to solving and classifying the solutions. Note that, throughout the discussion, we work in a gauge-fixed worldsheet metric with $\gamma_{(0)ab}=\eta_{ab}$. In general, the solution space is large and intricate enough, but we will try to go slowly through the full set. 
\subsubsection*{LO analysis}	
Using the LO Virasoro constraints in \eqref{lo virasoro}, we determine the gauge-fixed LO Virasoro constraints, now evaluated for static BTZ, as given below 
\begin{equation}\label{lo virasoro 1}    
T_{(0)\tau\tau} = T_{(0)\s\s}  = \frac{r_h^2}{2} \left[ \left( \partial_\t \xf \right)^2 + \left( \partial_\s \xf \right)^2 \right]=0\,,~~~T_{(0)\t\s}  = r_h^2\, \partial_\s \xf \cdot \partial_\t \xf  = 0\,.
\end{equation}    
The above equations \eqref{lo virasoro 1} imply that $x^\phi$ is independent of the worldsheet coordinates and hence trivial,
\begin{equation}\label{lo sol}
    x^\phi=x^\phi_0.
\end{equation}
The equations of motion in the gauge $\gamma_{(0)ab}=\eta_{ab}$ for the LO embedding field $x^\mu$ is
\begin{eqnarray}\label{lo eom}
    \frac{1}{2}\Gamma_{(0)}^{ab} \partial_ax^\nu \partial_b x^\lambda \partial_\mu \Omega_{\nu\lambda} &= \Gamma_{(0)}^{ab}\partial_a \left(\partial_b x^\nu \Omega_{\mu\nu} \right).
\end{eqnarray}
The solution from the LO Virasoro constraint equation is consistent with the equation of motion \eqref{lo eom}. The triviality of the solution is unsurprising, since the LO equation of motion equates a Lorentzian worldsheet geometry to the LO Euclidean target space geometry. Since this identification cannot be possible, the solution itself turns out to be trivial.
\medskip

The trivial solution at the LO \eqref{lo sol} can imply one of the two scenarios very close to the event horizon, either the string reduces to a point particle, or it folds onto itself like a straight rigid line and gets aligned along the radial direction \cite{Bars:1994sv,Bagchi:2023cfp}. The second scenario is compelling for two reasons. Firstly, strings in the tensionless limit, as found near horizons, have been shown to fold upon themselves \cite{Bagchi:2020ats, Bagchi:2021ban}. Secondly, this behaviour agrees with earlier studies which suggest that folded strings emerge near a black hole \cite{Bars:1994sv}. Similar observations, at the LO for a Schwarzschild black hole, were also made in \cite{Bagchi:2023cfp}. These observations suggest a universality in the behaviour of LO strings near a black hole. We postulate that this universal behaviour at the LO arises due to the emergence of a codimension two Euclidean geometry at the LO in the near-horizon expansion of any non-extremal black hole spacetime. Although the straight, rigid configuration appears more compelling, in our analysis we shall consider both possibilities - the string reducing to a point as well as the folded straight line configuration.

\subsubsection*{NLO analysis}\label{sec: nlo polyakov}	
Now we will analyse the NLO behaviour of the string to find first signs of non-triviality. Unlike the LO analysis, where only the LO target space metric contributed, the NLO analysis will take into account the NLO target space metric \eqref{static expansion 1} as well. The NLO target space metric has contributions from both the two-dimensional Rindler and the sphere ($S^1$). In our gauge, the NLO Virasoro constraints \eqref{nlo virasoro} take the following form after imposing \eqref{lo sol}
\begin{subequations}\label{nlo vir con}
\begin{align}
		T_{(2)\t\t} =T_{(2)\s\s} =\left( \partial_\t \xt \right)^2 + \left( \partial_\s \xt \right)^2- \left( \frac{\a}{\xr} \right)^2 \left( \left( \partial_\t \xr \right)^2 + \left( \partial_\s \xr \right)^2 \right) & = 0\,, \label{nlo virasoro 1}\\
		T_{(2)\t\s} = \partial_\t \xt \partial_\s \xt - \left( \frac{\a}{\xr} \right)^2\partial_\t \xr \partial_\s \xr & = 0 \,. \label{nlo virasoro 2}
\end{align}
\end{subequations}
The gauge fixed NLO Polyakov Lagrangian in the vicinity of a static BTZ black hole is  
\begin{multline}\label{nlo polyakov lagrangian static btz}
	\L_{_{P,\, NLO}} = \frac{T}{2} \Bigg[\frac{\a}{\xr} \left(\partial_\t \xr\right)^2 +\frac{\xr}{\a} \left(\partial_\s \xt \right)^2-\frac{\xr}{\a}\left(\partial_\t \xt \right)^2 - \frac{\a}{\xr} \left(\partial_\s \xr\right)^2 \\- 2r_h^2 \left(\partial_\s \xf \partial_\s \yf - \partial_\t \xf \partial_\t \yf \right) - 2r_h \xr \left( \left(\partial_\s \xf\right)^2 - \left(\partial_\t \xf\right)^2 \right) \Bigg].
\end{multline}
The equation of motion for $y^\phi$ has the same solution as the LO transverse embedding fields ($x^\phi$), that is, \eqref{lo sol}. Whereas the NLO Polyakov Lagrangian \eqref{nlo polyakov lagrangian static btz} yields the following equations of motion for $x^\phi$, $x^t$ and $x^\rho$ after imposing \eqref{lo sol}
\begin{subequations}\label{nlo eom}
\begin{align}
\partial_\t^2 \yf - \partial_\s^2 \yf&=0\,,\label{nlo eom 1}\\
( \partial_\t^2 - \partial_\s^2 )\xt + \frac{1}{\xr} \left( \partial_\t \xt \partial_\t \xr - \partial_\s \xt \partial_\s \xr \right)&=0\,,\label{nlo eom 2}\\
\partial_\t \left( \frac{\partial_\t \xr }{\xr} \right) - \partial_\s \left( \frac{\partial_\s \xr }{\xr} \right) + \frac{1}{2\a^2}\left[ \left(\partial_\t \xt \right)^2 - \left(\partial_\s \xt \right)^2 \right] \nn \\+\frac{1}{2 \left( \xr \right)^2 } \left[ \left(\partial_\t \xr \right)^2 - \left(\partial_\s \xr \right)^2 \right]&=0\label{nlo eom 3}\,.
\end{align}
\end{subequations}
The NLO transverse embedding scalars $y^\phi$ follow the 2D wave equation. Thus, at the NLO, the string retains its usual excitation modes along the transverse direction. The equations of motion for the LO longitudinal embedding fields ($x^t, x^\rho$) given in \eqref{nlo eom 2} and \eqref{nlo eom 3} and the Virasoro constraints \eqref{nlo vir con} give rise to multiple classes of solutions. In what follows, we will try to assume classes of worldsheet separable ans\"atze for such solutions, which lets us explore the parameter space with considerable generality. 

\subsection{Classifying Polyakov solutions: Static BTZ}

We can arrive at a general set of solutions by considering two ans\"atze for the embedding fields\footnote{Note that $x^\phi=$ constant is a solution of the LO magnetic theory.} $\xt$ and $\xr$. The first of them being
\begin{equation}\label{ansatz1}
    \xt(\tau,\sigma)=f(\tau)\,,~~
    \xr(\tau,\sigma)=g(\tau)h(\sigma)\,,
\end{equation}
with $f,g,h$ being arbitrary functions.
From the structure of the above ansatz, it follows that the NLO Virasoro constraints \eqref{nlo vir con} can be solved by the method of separation of variables. For above ansatz \eqref{ansatz1}, the NLO Virasoro constraints \eqref{nlo vir con} then take the form 
\begin{subequations}
\begin{align}\label{eq:separation of variables ansatz1}
\dot f(\t)^2 - \frac{\a^2}{g(\t)^2}\dot g(\t)^2 = \frac{\a^2}{h(\s)^2}h'(\s)^2\, &=\vartheta^2, \\
\frac{\dot g(\tau) h'(\sigma)}{g(\tau)h(\sigma)} &= 0 \, , 
\end{align}
\end{subequations}

where $\vartheta$ is the separation constant. For $\vartheta = 0$, we get
\begin{equation}\label{static rindler geodesic}
    \xt = \xt_0 \pm \frac{\ell^2}{2r_h} \ln \left( \frac{\xr}{\xr_0} \right),~~~x^\rho=x^\rho_0\tau.
\end{equation}

Here, we use  $\alpha = \frac{\ell^2}{2r_h}$. The expression for $x^\rho$ is obtained using the equation of motion of $x^t$ in \eqref{nlo eom 2}. The above solution \eqref{static rindler geodesic} also satisfies \eqref{nlo eom 2}, \eqref{nlo eom 3}, and describes a null geodesic in Rindler spacetime. This can be verified by solving the relevant equation: $\t_{\m\n}\frac{dx^\m}{d\lambda}\frac{dx^\n}{d\lambda} = 0$, where $\lambda$ is an affine parameter that parametrises the null geodesic. The solutions of the LO embedding fields and the NLO transverse embedding field paint a physical picture wherein, for an asymptotic observer, the string seems to shrink to a point particle, then follow null geodesics while retaining its stringy nature at the NLO transverse directions\footnote{Note that there is no $\sigma$ periodicity in this solution, hinting at pure particle nature of the LO solution.}.
\medskip

For $\vartheta \neq 0 $, 
the solution to the equations of motion of the LO embedding coordinates and the NLO Virasoro constraint is given as follows
\begin{equation}\label{yoyo strings}
x^\phi=x^\phi_0\,,~~~\xt (\t,\s) = \xt_0 \pm \vartheta\t \,,~~~\xr(\t,\s)= \xr_0 \exp \left( \pm \frac{\vartheta}{\alpha} \sigma \right)\,.
\end{equation}

\begin{figure}[hbt!]
	\centering
	\includegraphics[width=0.95\linewidth]{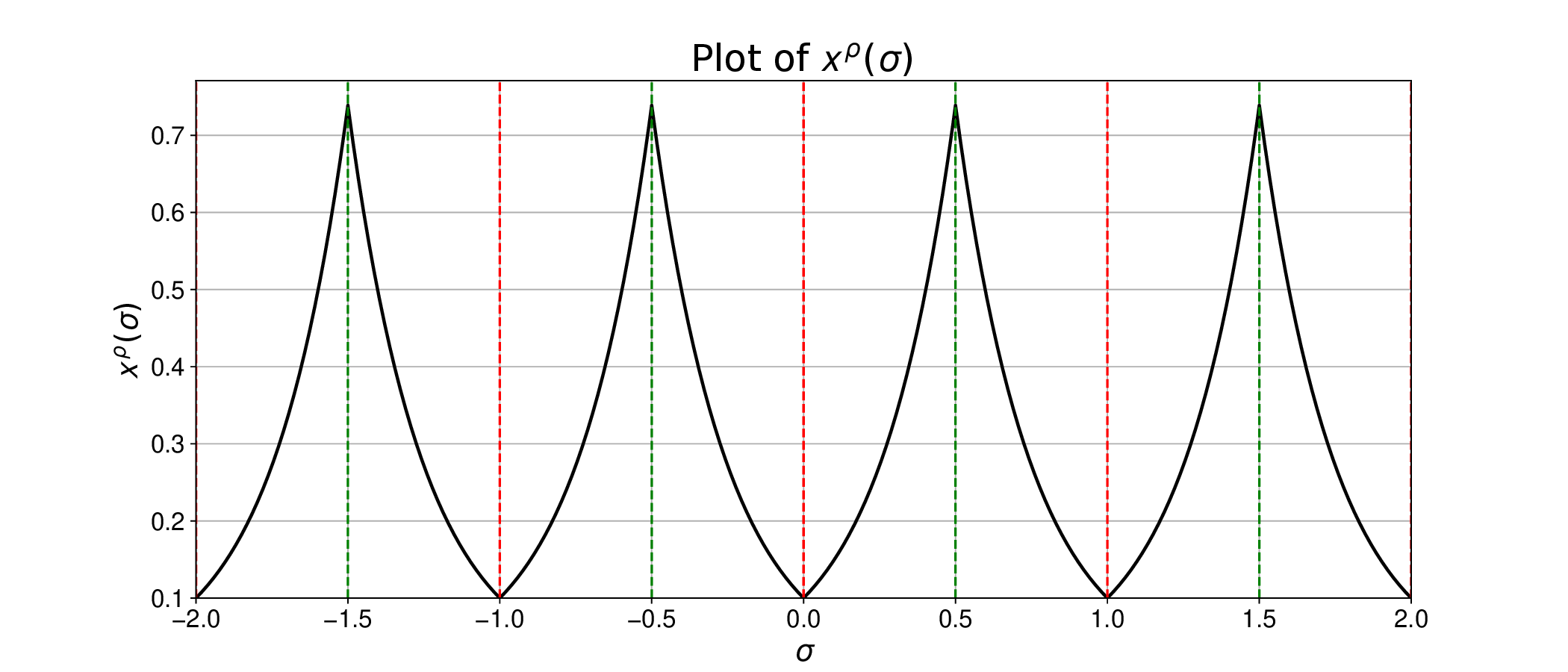}
	\caption{This is a plot for $\xr(\s)$ vs $\s$ in \eqref{yoyo strings 2} depicting ``yo-yo" strings. The constants have been chosen to be $\xr_0 = 0.1$, $\vartheta = 4$ and $\alpha = 1$.}
    \label{fig: yoyo strings}
\end{figure}

Observe that the solution for $x^\rho$ is static in time but not continuous, 
given that $\sigma$ is a compact worldsheet coordinate, say in the interval $[0,1)$. However, using \eqref{yoyo strings}, we can construct a static ``yo-yo" string solution \cite{Bars:1994sv,PhysRevD.13.2364} depicted in figure \ref{fig: yoyo strings}, that is continuous but has isolated non-differentiable points, as follows 
\begin{equation}\label{yoyo strings 2}
			x^\rho(\s) = 
			\begin{cases}
				\xr_0 \exp \left( \frac{\vartheta}{\alpha} \sigma \right)~~~~~~\qquad\s \in \left[ 0,\frac{1}{2}\right]\\
				\xr_0 \exp \left( \frac{\vartheta}{\alpha} (1-\s) \right)~~\quad\s \in \left[\frac{1}{2},1 \right]\,.
			\end{cases}
\end{equation}

\begin{figure}[hbt!] 
		\centering
		\includegraphics[width=0.92\linewidth]{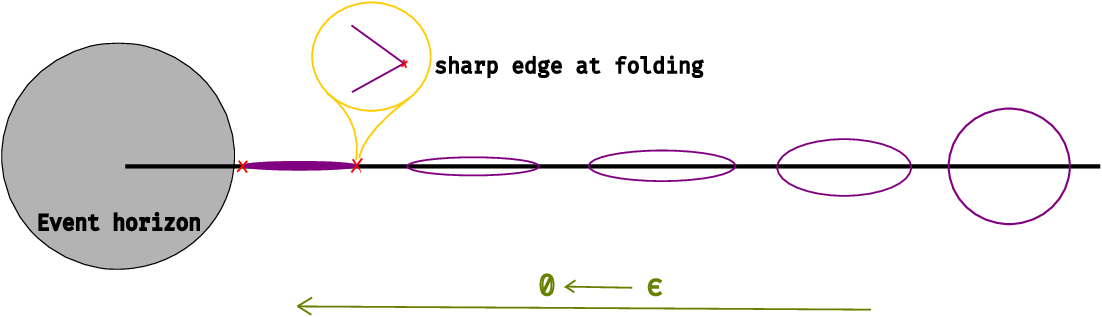}
		\vspace*{-5mm}
		\caption{Exaggerated depiction of the string folding onto itself radially and approaching the blackhole along $x^\phi=0$.}
        \label{fig:folded strings}
	\end{figure}

Let us ruminate on the mathematical consequences of the yo-yo string solution \eqref{yoyo strings 2}. If we assume \eqref{yoyo strings 2} to be the correct solution of the radial LO embedding scalars, the worldsheet manifold is no longer smooth, specifically at $\s=0,\frac{1}{2}$. This is reminiscent of folded strings, which at the LO tend to shrink their angular spread \eqref{lo sol} and align along the radial direction. An exaggerated diagram of this behaviour is given in figure \ref{fig:folded strings}. The yo-yo string solution seems to reinforce this conclusion, while retaining stringy excitation modes at the NLO \eqref{nlo eom 1}. We can think that in the course of its journey towards the event horizon, the erstwhile closed string gets folded. As a result, the ``bends" (or special points) at the two radial extremities become sharp\footnote{This derivative discontinuity appearing on the pullback fields can remind the learned reader of ``spiky strings" \cite{Gubser:2002tv, Kruczenski:2004wg, Jevicki:2008mm} which mostly appear in AdS ``long string'' limits. Our solution is like a two-spike solution made out of pinching a closed circular string at two opposite points.}. Very close to the horizon, discarding the NLO wavy fields, the bends become abruptly sharp (as shown in figure \ref{fig: yoyo strings}), thus destroying the differentiable structure of the worldsheet at the bends. To an asymptotic observer, the string would appear to be like a rigid ``stick" at rest, very close to the event horizon (cf. Figure \ref{fig:folded strings}). 

Now we move on to consider the second ansatz, which offers more wiggle room to the solution space than the previous one, and has the following form 
\begin{eqnarray}\label{ansatz2}
\xt(\tau,\sigma)=f(\tau)+\psi(\sigma)\,,~~
    \xr(\tau,\sigma)=g(\tau)h(\sigma)\,.
\end{eqnarray}
Now one can see we have moved beyond the stationary string regime as coordinate time also includes $\sigma$ dependent factors.
This ansatz also allows us to solve the NLO Virasoro constraints \eqref{nlo vir con} using the separation of variables method. Substituting the ansatz, the constraints take the form 
\begin{subequations}\label{sol gen ansatz}
\begin{align}
A(\t)+ B(\s) &= 0 \,, \label{sol gen ansatz 1}\\
\left(\frac{g(\t)}{\a} \right) \frac{\dot f(\t)}{\dot g(\t)} = \left(\frac{\a}{h(\s)} \right) \frac{h'(\s)}{\psi'(\s)} &= \theta \,, \label{sol gen ansatz 2}
\end{align}
\end{subequations}
where \(\theta\) is a constant of separation that determines the features of the solution and \(A(\t)\), \(B(\s)\) are defined as 
\begin{eqnarray}
 A(\t)= \dot f(\tau)^2 - \frac{\a^2}{g(\t)^2}\dot g(\t)^2\,,~~~B(\s)=\psi'(\s)^2 - \frac{\a^2}{h(\s)^2}h'(\s)^2\,.
\end{eqnarray}
Being functions of different variables $A(\t)$ and $B(\s)$ should necessarily be a constant, say $\omega^2$, with opposite signs. Here, our solutions bifurcate into two cases $\omega^2=0$ and $\omega^2\neq0$, on top of the $\theta$ dependent classification. We will consider all the cases below and summarise later on.
\medskip

\ding{112} \uline{\textbf{Classification based on $\omega$}}\\

\textbf{\emph{$\blacksquare$ Case $\omega = 0$:}} In this case, equations \eqref{sol gen ansatz} imply
\begin{equation}\label{eq: omega 0 virasoro}
\theta =
    \begin{cases}
1&\implies f(\t) =\alpha \ln g(\t)\,, ~~~\psi(\s) = \alpha \ln h(\s)\,,\\
-1&\implies f(\t) =-\alpha \ln g(\t)\,, ~~~\psi(\s) = -\alpha \ln h(\s)\,.
    \end{cases}
\end{equation}
In this scenario\footnote{The converse, that is, $\theta=\pm1\implies\omega^2=0$ is also true.}, the LO embedding fields take the form
\begin{equation}\label{eq: omega 0 nlo rindler}
    \xf = \xf_0 \,, ~~~\xt(\t,\s) =\xt_0 \pm \alpha \ln \lb \frac{\xr(\t,\s)}{\xr_0} \rb\,, ~~~ \xr(\t,\s) = \xr_0~g(\t)h(\s) \,.
\end{equation}

The equation \eqref{eq: omega 0 nlo rindler} is that of a null Rindler geodesic in two dimensions, but with an implicit $\sigma$ dependence that points towards a continuous family of null geodesics, one sitting at each spatial point on the string. To determine the exact functional form of $f(\t)$ and $\psi(\s)$, one has to invoke the NLO equations of motion \eqref{nlo eom 2} and \eqref{nlo eom 3}, which, using the ansatz \eqref{ansatz1}, take the form\footnote{The $\pm$ in \eqref{eq: omega 0 nlo eom} refer to the $\theta=\pm1$ cases respectively.}
\begin{subequations}\label{eq: omega 0 nlo eom}
    \begin{align}
        \ddot f(\t) \pm \frac{\dot f(\t)^2}{\alpha} - \psi''(\s) \mp \frac{\psi'(\s)^2}{\alpha}&=0 \,, \label{eq: omega 0 nlo eom 1}\\
        \pm \ddot f(\t) + \frac{\dot f(\t)^2}{\alpha} \mp \psi''(\s) - \frac{\psi'(\s)^2}{\alpha}&=0 \,. \label{eq: omega 0 nlo eom 2}
    \end{align}
\end{subequations}

We have obtained the equation for $f(\t)$ and $\psi(\s)$ in a variable separable form. If we choose the $\boldsymbol +$ sign and set the separation constant of \eqref{eq: omega 0 nlo eom 1} to $\pm K^2$, then the separation constant of \eqref{eq: omega 0 nlo eom 2} automatically becomes $\pm K^2$. However, if we choose the $\boldsymbol{-}$ sign and set the separation constant of \eqref{eq: omega 0 nlo eom 1} to be $\pm K^2$, then the separation constant of \eqref{eq: omega 0 nlo eom 2} becomes $\mp K^2$. The functional forms of $f(\t)$ and $\psi(\s)$ are presented below, categorised by case, while the functions $g(\t)$ and $h(\s)$ can be obtained using equation \eqref{eq: omega 0 virasoro}\footnote{Note that the cases are with respect to the separation constant of \eqref{eq: omega 0 nlo eom 1}. In all instances, we have taken $K>0$.}.\\

\textbf{\emph{$\diamondsuit$ Subcase: $K = 0$:}} Here, we get the following functional forms for $f(\t)$ and $\psi(\s)$
\begin{equation}\label{eq:K=0}
    f(\t) = f_0 \pm \alpha \ln \lb \t - \t_0 \rb\,,~~~\psi(\s) = \psi_0 \pm \alpha \ln \lb \s - \s_0 \rb\,.
\end{equation}
This is a scaling solution where the radial target coordinate collapses to zero at finite values of worldsheet coordinates. This means that near the special point $(\sigma_0,\tau_0)$, different points on the string are focused on the same radial spacetime point.\\

\textbf{\emph{$\diamondsuit$ Subcase: $K^2$:}} For this choice of the separation constant, we get the following forms for $f(\t)$ and $\psi(\s)$
\begin{equation}\label{eq:+K}
    f(\t) = f_0 \pm \alpha  \ln \left(\cosh \left(\frac{K (\t - \t_0)}{\sqrt{\alpha }}\right)\right)\,,~~~\psi(\s) = \psi_0 \pm \alpha  \ln \left(\cosh \left(\frac{K (\s - \s_0)}{\sqrt{\alpha }}\right)\right)\,.
\end{equation}
This is more like a `kink' or soliton-like solution which is non-singular everywhere on the worldsheet. One can think of this as an excitation in the radial direction, centering on the special point $(\sigma_0,\tau_0)$ where the worldsheet reaches its closest approach to the horizon.\\

\textbf{\emph{$\diamondsuit$ Subcase: $-K^2$:}} In this case, the functions $f(\t)$ and $\psi(\s)$ take the following form 
\begin{equation}\label{eq:-K}
    f(\t)= f_0 \pm \alpha  \ln \left|\cos \left(\frac{K (\t - \t_0)}{\sqrt{\alpha }}\right)\right|\,,~~~\psi(\s) = \psi_0 \pm \alpha  \ln \left| \cos \left(\frac{K (\s - \s_0)}{\sqrt{\alpha }}\right)\right| \,.
\end{equation}
Here, the radial embedding is periodic and touches $x^\rho=0$ at intervals. One would think of this as a solution that periodically (in $\sigma$) collapses to the horizon. The modulus makes sure we always have a positive radial distance.
\medskip

This concludes our discussion on the case $\omega=0$. Note again that even though the LO embedding fields seemingly take the form of a two-dimensional null Rindler geodesic \eqref{eq: omega 0 nlo rindler}, they still depend on the worldsheet $\sigma$ coordinate, indicating the memory of their stringy origin. This is in contrast to the first ansatz \eqref{ansatz1}, where we obtained the null Rindler solution \eqref{static rindler geodesic}, but the embedding fields depended solely on the $\tau$ coordinate. Next, we will discuss the $\omega^2\neq0$ case to see if such non-triviality persists.\\

\textbf{\emph{$\blacksquare$ Case $\omega > 0$:}} In this case, the solutions to the equations of motion of the LO embedding coordinates and the NLO Virasoro constraint are given as
\begin{equation}\label{eq: magnetic NLO theta solution}
\xf=\xf_0,~~\xt (\t,\s) = \xt_0 + \b (\theta\t\pm\s ),~~\xr (\t,\s)= \xr_0 \exp \left[\textcolor{red}{\pm}\frac{\b}{\a} \left( \t\pm\theta \s \right)\right],
\end{equation}
where $\beta = \frac{\omega}{ \sqrt{|\theta^2-1|} }$ and $|\theta| \neq 1$. The $\pm$ signs in front of $\sigma$ are to ensure periodicity of the $\sigma$ worldsheet coordinate. The $\pm$ sign ensures that a periodic string solution can be constructed by gluing two solutions as was done to construct yo-yo strings in \eqref{yoyo strings 2}, making the above a generalisation of those folded strings. However, the $\pm$ sign on the overall exponential in $\xr$ solution characterises the nature of the solution \eqref{eq: magnetic NLO theta solution}. We mark this sign in red to distinguish it from other sign choices in this section.  
\medskip

These solutions are interesting in the sense that they can be thought of worldsheet `boosted' version of Rindler geodesics, with each constant $\sigma$ point following a boosted Rindler trajectory\footnote{What we mean here becomes clear by considering two different points on the string: $\sigma = 0$ and $\sigma = 1$.
For $\sigma = 0$: $x^t= x^t_0 + \beta\theta\tau,\quad 
x^{\rho} = x^{\rho}_0e^{-(\beta/\alpha)\tau}$
For $\sigma = 1$: $
x^t = x^t_0 + \beta\theta\tau + \beta,\quad 
x^{\rho} = x^{\rho}_0 e^{-(\beta/\alpha)(\tau + \theta)}$. So these are still a family of Rindler trajectories starting at different initial times and different initial radial positions, but same proper acceleration.}.
Here, the boost parameter is $\theta$ that mixes worldsheet space and time coordinates and, in principle, should vary between $0$ and $1$, with the latter resulting in the emergence of a null worldsheet, which our solution space does not admit. In what follows, we analyse each solution generated by the ansatz \eqref{ansatz2} on a case-by-case basis for values of $\theta$. Note that, hereafter, in all the cases where a sign is mentioned, we refer to the ones in red as evident in \eqref{eq: magnetic NLO theta solution}.\\

\ding{112} \uline{\textbf{Classification based on $\theta$}}\\

\textbf{\emph{$\blacksquare$} \emph{Case $\theta=0$:}} The solutions for LO embedding fields take the form
\begin{eqnarray}\label{case2}
\xf=\xf_0,~~~\xt (\s) = \xt_0 \pm \omega  \s,~~~\xr (\t)= \xr_0 \exp \left( \pm \frac{\omega\t}{\a}  \right).
\end{eqnarray}
Notice that $x^t(\s)$ is linear in the compact direction $\sigma$, implying that the LO timelike embedding coordinate becomes periodic, which indicates the presence of a situation akin to closed timelike curves (CTC). CTCs are fatal pathologies in Lorentzian theories. To avoid these pathologies, one needs to set $\omega=0$. However, setting $\omega=0$ in \eqref{case2} reduces the solution to a static point in the target spacetime manifold.\\

\textbf{\emph{$\blacksquare$}\emph{Case $|\theta|\not\in \{0,1\}$ and \textcolor{red}{$+$}:}} 
In this case, we recover folded strings, but now these folded strings move radially outwards, while elongating at the same time. The length of the folded string increases exponentially as it moves outwards.\\
        
\textbf{\emph{$\blacksquare$}\emph{Case $|\theta|\not\in \{0,1\}$ and \textcolor{red}{$-$}:}} The solutions which are obtained in this case represent folded strings falling radially inwards. The length of the folded string decreases exponentially as it moves inwards. As it happens, we can explicitly see that the string becomes more like a point as it approaches the event horizon. For this, we will establish an asymptotic relationship between $\xt$ and $\xr$. First, note that the negative branch of \eqref{eq: magnetic NLO theta solution} can be manipulated to get the following relationship
\begin{equation}\label{eq: pre-asymptotic}
    x^t - \xt_0 =  - \alpha\theta\ln \lb \frac{\xr}{\xr_0} \rb + \beta (s_2 \theta^2 + s_1)\sigma \,,
\end{equation}

where, $s_1$ and $s_2$ can independently take the values $\pm 1$. In this equation, we would like to take the asymptotic limit in the target space $\xt(\t,\s) - \xt_0 \gg 0$, which corresponds to the late worldsheet time limit $\tau \gg 0$. In this regime, we also get $\xr / \xr_0 \ll 1$. Now we consider the quantity
\begin{equation}\label{eq: asymptote 1}
    \frac{\xt - \xt_0}{- \alpha\theta\ln \lb \frac{\xr}{\xr_0} \rb} = 1 + \frac{ \beta (s_2 \theta^2 + s_1)\sigma }{ - \alpha\theta\ln \lb \frac{\xr}{\xr_0} \rb  } \,.
\end{equation}

At this point, we should remark that $\sigma$ is a compact coordinate, which implies that  $\beta (s_2 \theta^2 + s_1) \sigma$ is bounded. So, in our asymptotic limit, the second term of the RHS of \eqref{eq: asymptote 1} dies off, and thus, we get the following relation in this limit
\begin{equation}\label{mimic rindler}
\xt \sim \xt_0 - \frac{\theta \ell^2}{2r_h} \ln \left( \frac{\xr}{\xr_0} \right)\,.
\end{equation}

\begin{figure}[hbt!]
		\centering
		\includegraphics[width = 0.93\linewidth]{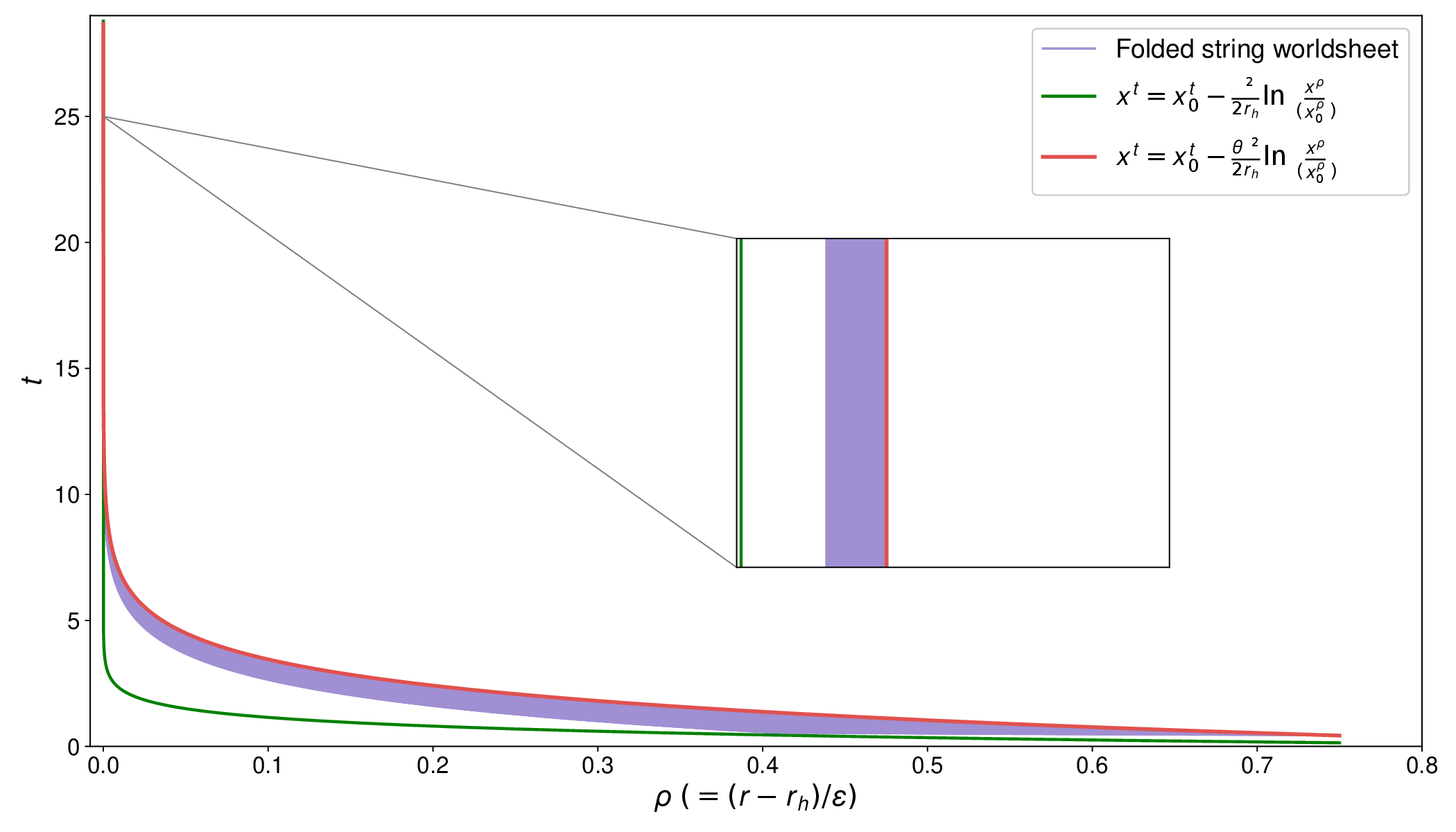}
		\caption{This figure depicts the behaviour of the folded string worldsheet as it falls towards the event horizon of a BTZ black hole. The string tends to mimic the null Rindler geodesic as it approaches the event horizon. In this plot, we have taken $\xt_0 = 0$, $\xr_0 = 1$, $\alpha = \ell^2/2r_h=0.5$, $\beta = 0.2$ and $\theta = 3$. %The inset plot zooms into the region $[0,5\times10^{-7}]\times[7,29]$ in the $\rho - t$ plane. 
        It can be noticed in the zoomed inset plot that the shrinking string asymptotes to \eqref{mimic rindler}, rather than the pure null Rindler geodesic \eqref{static rindler geodesic}.}
        \label{fig: Rindler dynamics}
\end{figure}

\begin{figure}[hbt!]
    \centering
    \includegraphics[width=0.995\linewidth]{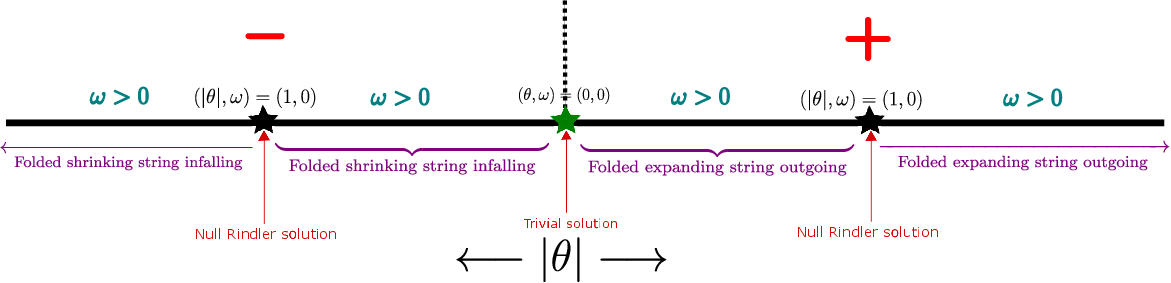}
    \caption{The full parameter space of magnetic strings with different classes of solutions marked.}
    \label{fig: magnetic string parameter space}
\end{figure}

This equation tells us that the infalling folded string tends to ``mimic'' an ingoing null Rindler geodesic as it approaches the horizon. Moreover, the form of \eqref{mimic rindler} tells us that the string completely looses its radial width while approaching the event horizon of a static BTZ black hole. Notice that the parameter $\theta$, as described before, signifies the deviation from the null Rindler geodesic. This motion of the folded string falling towards the black hole while shrinking is depicted in figure \ref{fig: Rindler dynamics}. The purple region denotes the worldsheet of a folded string as it falls towards the event horizon located at $\rho=0$. The green curve represents a null geodesic in Rindler spacetime and the red curve shows \eqref{mimic rindler}. Figure \ref{fig: Rindler dynamics} depicts the embedded worldsheet in the target spacetime. The full parameter space of magnetic strings as foliated by values of $\omega$ and $\theta$ is depicted in Figure \ref{fig: magnetic string parameter space}.
\medskip


 This brings an end to our expansive discussion on static BTZ strings. All this information together could be a bit daunting to the reader, so the solutions analysed above in the vicinity of a static BTZ black hole are summarised in the table \ref{tab:solutions} for a handy reference. To provide a mid-section broad summary, we could just say that LO strings near BTZ horizon either become (families of) null geodesics or folded strings which shrink/grow as they fall in/go away from the black hole.

\subsection*{An aside on worldsheet tidal forces}
Our discussions above had a crucial subtlety, folded strings falling into the horizon seemed to shrink with coordinate time.
However, the reader should be assured that this string solution does not go against our intuitions of tidal effects in general relativity, where one expects the string to expand as it approaches the horizon. This can be concretely seen from the geodesic deviation equation for a one-parameter family of geodesics $X^\mu(\t,q)$
\begin{equation}
    \frac{D^2 S^\m}{d\tau^2} = -R^\m_{\n\rho\lambda}\frac{d X^\n}{d\t} S^\rho \frac{d X^\lambda}{d\t}\,,
\end{equation}
where
\begin{equation}
    S^\m(\tau) = \left. \frac{\partial X^\mu}{\partial q} \right|_{q=0}\,,
\end{equation}
is the deviation vector on the geodesic $X^\m(\t)\equiv X^\m(\t,0)$, where $\t$ is an affine parameter, $R^\m_{\n\rho\lambda}$ is the Riemann curvature tensor of target space evaluated on the geodesic and 

$D/d\t = \frac{dX^\m}{d\t}\nabla_\m$ is proportional to the covariant derivative. We will denote $\dot X^\mu = dX^\mu/d\tau$ in what follows. For radial geodesics $(\dot X^\phi = 0)$ in a static BTZ black hole in Boyer-Lindquist-like coordinates \eqref{general btz}, the relevant deviation vector can be written in the form $S = S^t\partial_t + S^r\partial_r$. In this case, the geodesic deviation equation in the full non-expanded background becomes
\begin{subequations}\label{eq: geodesic deviation static btz}
    \begin{align}
        \frac{D^2 S^t}{d\t^2} &= -R^t_{trr}\dot X^t \dot X^r S^r - R^t_{rtr}\lb \dot X^r \rb^2 S^t \nn\\
        &= \frac{1 }{(X^r)^2-r_h^2}\left( S^t \dot X^r  - S^r \dot X^t  \right)\dot X^r  \,, \label{eq: geodesic deviation static btz 1}\\
        \frac{D^2 S^r}{d\t^2} &= -R^r_{rtt}\dot X^t \dot X^r S^t - R^r_{trt}\lb \dot X^t  \rb^2 S^r \nn \\
        &=  \frac{\left( (X^r)^2 - r_h^2 \right) }{ \ell^4} \left( S^t \dot X^r - S^r \dot X^t  \right)\dot X^t \,. \label{eq: geodesic deviation static btz 2}
    \end{align}
\end{subequations}

As we are working in the NH regime, we would like to impose the same on the above \eqref{eq: geodesic deviation static btz}. $X^\m(\t)$ will take the usual NH expansions \eqref{X expansion}, which in this coordinate system becomes
\begin{equation}\label{eq: x expansion}
    X^t(\t) = x^t(\t) + \epsilon y^t(\t) +\O(\epsilon^2)\,, ~~~~~~~~ X^r(\t) = r_h + \epsilon x^\rho (\t) + \O (\epsilon^2) \,.
\end{equation}

Similarly, the deviation vector $S^\m(\t)$ will also admit an NH expansion as follows
\begin{equation}\label{eq: s expansion}
    S^t(\t) = s^t(\t) + \epsilon \xi^t (\t) + \O(\epsilon^2)\,, ~~~~~~ S^r(\t) = \epsilon s^\rho(\t) + \O(\epsilon^2) \,.
\end{equation}

In this setting, the geodesic deviation \eqref{eq: geodesic deviation static btz} expands as
\begin{subequations}\label{eq: expanded geodesic deviation static btz}
    \begin{align}
        \left. \frac{D^2 S^t}{d\t^2}\right|_{LO} + \O(\epsilon) &= \lb \frac{  \dot x^\rho \left(s^t  \dot x^\rho  - s^\rho \dot x^t \right)}{2 r_h x^\rho  }\rb \epsilon + \O(\epsilon^2) \,, \label{eq: expanded geodesic deviation static btz 1}\\
        \left. \frac{D^2 S^r}{d\t^2}\right|_{LO} \epsilon + \O(\epsilon^2) &= \lb \frac{x^\rho \dot x^t \left( s^t \dot x^\rho  - s^\rho  \dot x^t\right)}{2 \alpha ^2 r_h}\rb \epsilon^2 + \O(\epsilon^3) \,, \label{eq: expanded geodesic deviation static btz 2}
    \end{align}
\end{subequations}

where,
\begin{subequations}\label{forces2}
    \begin{align}
        \left. \frac{D^2 S^t}{d\t^2}\right|_{LO} &= \ddot s^t +\frac{1}{4} \left(\frac{2 \left(2 \dot s^\rho\dot x^t + s^\rho\ddot x^t + 2 \dot s^t \dot x^\rho + s^t \ddot x^\rho \right)}{x^\rho} - \frac{\dot x^\rho \left(2 x^\rho \dot x^t + s^t \dot x^\rho \right)}{\lb x^\rho\rb^2} + \frac{s^t \lb \dot x^t \rb^2}{\alpha ^2}\right) , \\
        \left. \frac{D^2 S^r}{d\t^2}\right|_{LO} &= \ddot s^\rho  + \frac{1}{4 \alpha ^2 \lb x^\rho \rb^2}\left(  \lb x^\rho \rb^2 \left( s^\rho \lb\dot x^t\rb^2 + 2 s^t \dot x^\rho \dot x^t \right) - 2 \alpha ^2 x^\rho \left(2 \dot s^\rho \dot x^\rho + s^\rho \ddot x^\rho \right)\right.  \nn \\
        & \quad \left. + 3 \alpha ^2 s^\rho \lb \dot x^\rho \rb^2 + 2 \lb x^\rho\rb^3 \left(2 \dot s^t \dot x^t + s^t \ddot x^t\right)  \right) .
    \end{align}
\end{subequations}

The expanded geodesic deviation equation \eqref{eq: expanded geodesic deviation static btz} show that the LO dynamics is unaffected by tidal effects appearing at different orders of $\epsilon$, which implies that LO string dynamics do not observe any tidal forces. In fact, the string, having a Lorentzian worldsheet $\gamma_{(0)ab}$, is compelled to shrink to a point as it reaches the event horizon since the horizon metric $\Omega_{\m\n}$ is of Euclidean signature, which means any Lorentzian embedding on it necessarily has to be trivial. This explains the shrinking of the NH magnetic string as it goes towards event horizon. 
\medskip

The leading order LHS in \eqref{eq: expanded geodesic deviation static btz}, given by \eqref{forces2}, is actually the LHS of the geodesic deviation equation in 2D Rindler spacetime $ds^2 = -\frac{\rho}{\alpha} dt^2 + \frac{\alpha}{\rho}d\rho^2$. This might tempt the reader to conclude that the near-horizon regime is flat. However, we see the non-triviality of the NH region when we introduce a small perturbation $\delta X^\phi(\tau)$ to our radial geodesics. This perturbation introduces extra $\epsilon$ expansions in \cref{eq: x expansion,eq: s expansion} respectively which in turn introduces $\delta x^\phi$ and $\delta s^\phi$ corrections in \eqref{eq: expanded geodesic deviation static btz} at commensurate $\epsilon$ orders, as opposed to the purely radial case. 
\medskip

This hints to us that even an ``almost radial" geodesic family will face tidal drag in the radial direction. Only when the geodesic is exactly radial does the tidal drag vanish, which is the case for our magnetic string solutions. It should be further noted that in the full geometry at asymptotic distances, the string may have experienced tidal forces, but only as it moves towards the NH region, such forces start occurring at different orders, leading to our situation.

\medskip

\newcolumntype{M}[1]{>{\centering\arraybackslash}m{#1}}
\begin{table}[hbt!]
    \centering
    \begin{tabular}{|m{2.5cm}|M{3cm}|M{2cm}|M{2cm}|M{4cm}|}
    \hline
    \centering Ans\"atze &$\xt(\tau,\sigma)$ & $\xr(\tau,\sigma)$ & Condition on parameters & Physical Interpretation\\
    \hline
    \multirow{2}{*}{\parbox[c][3\baselineskip][c]{2.5cm}{\centering Ansatz \eqref{ansatz1}}}&$\xt_0+\alpha \ln(\tau-\tau_0)$ & $\xr_0(\tau-\tau_0)$ & $\vartheta=0$ & The string shrinks to a point and follows null Rindler geodesics.\\
    \cline{2-5}
    &$\xt_0\pm\vartheta\tau$&$\xr_0e^{\pm\frac{\vartheta}{\alpha}\sigma}$&$\vartheta\neq0$& Folded string with constant length.\\
    \Xhline{3\arrayrulewidth}
    \multirow{4}{*}{\parbox[c][13\baselineskip][c]{2.5cm}{\centering Ansatz \eqref{ansatz2}}}&$\xt_0$ & $\xr_0$ & $\theta=0$ and $\omega=0$ & Trivial solution.\\
    \cline{2-5}
    &$\xt_0 \pm \alpha \ln \lb \frac{\xr(\t,\s)}{\xr_0} \rb$ & $\xr_0~g(\t)h(\s)$& $\theta=\pm1$ and $\omega=0$ & The string follows a family of null Rindler geodesics\footnotemark.\\
    \cline{2-5}
    &$\xt_0+\beta(\theta~\tau\pm\sigma)$&$\xr_0e^{\frac{\beta}{\alpha}\left(\tau\pm\theta\sigma\right)}$&$|\theta|\neq 0,1$, $\omega>0$ and \textcolor{red}{$+$} (outgoing)& Folded string with increasing length with coordinate time. But the string moves away from the black hole. \\
    \cline{2-5}
    &$\xt_0+ \beta(\theta~\tau\pm\sigma)$&$\xr_0e^{-\frac{\beta}{\alpha}\left(\tau\mp\theta\sigma\right)}$&$|\theta|\neq 0,1$, $\omega>0$ and \textcolor{red}{$-$} (infalling)& Folded string with decreasing length with coordinate time and asymptotically traces null Rindler geodesic\\
    \hline
    \end{tabular}
    \caption{This table summarises all the solutions for the LO embedding fields in the magnetic case.} 
    \label{tab:solutions}
\end{table}
\footnotetext{{The functional form of the functions $g(\t)$ and $h(\s)$ can be deduced from \cref{eq: omega 0 virasoro,eq:K=0,eq:+K,eq:-K}.}}

\subsection{Polyakov strings near BTZ black hole: Rotating case}
In this section, we study strings near a rotating BTZ black hole using the string-Carroll expansion. The near-horizon metric of a rotating BTZ black hole, expressed in the co-rotating frame \eqref{non extremal expansion components}, exactly matches the near-horizon metric of a static BTZ black hole \eqref{static expansion components}, upon making the replacements $x^t\to x^T$, $r_h\to r_+$, and $\alpha\to \alpha_+$. The slight modification of the components of $\Theta_{\mu\nu}$ does not affect the dynamics of the string. Consequently, the string dynamics near a rotating BTZ black hole in the co-rotating frame are equivalent to those in the static case. In what follows, we will remark on the characteristics of the string in the non-co-rotating frame \footnote{We will work in the same gauge as in the static case, ie, $\Gamma^{ab}_{(0)}=\eta^{ab}$.}. 
\medskip

The map of the near-horizon expansion of a rotating BTZ black hole in the non-co-rotating frame and the string-Carroll expansion is given as 
\begin{subequations}\label{eq:rotating BTZ black hole map}
    \begin{align}
        \Omega_{\mu\nu}dx^\mu dx^\nu&=~r_+^2 \left(d\phi-\frac{J}{2 r_+^2}dt\right)^2\,,\\
        \lb\tau_{\mu\nu}+\Theta_{\mu\nu}\rb dx^\mu dx^\nu&=~-\frac{2r_+ \rho}{\ell^2} dt^2 + \frac{\alpha_+}{\rho} d\rho^2+2r_+\rho \, d\phi^2\,.
    \end{align}
\end{subequations}

The Virasoro constraints and the equation of motion for the LO transverse embedding field $x^\phi$ in the non-corotating frame are \eqref{lo virasoro} and \eqref{lo eom} respectively, which are solved by
\begin{align}\label{ne lo sol}
		%\xff (\t,\s)  &= \xff_0\, , \quad \text{(constant)} \nn\\
		\xf(\t,\s) & = \xf_0 + \frac{J}{2r_+^2}\xt(\t,\s).
\end{align} 
Here, the LO equation of motion and the Virasoro constraints do not solve for $x^t(\t,\s)$. This is an effect of the non-co-rotating frame that mixes the $\phi$ and $t$ coordinates. However, $x^\phi$ is constant for constant $x^t$, implying that the string shows a point or stick-like behaviour.

\subsubsection*{NLO analysis}\label{sec: nlo polyakov ne}
The NLO Virasoro constraints are as follows
\begin{subequations}\label{eq:NRC NLO Vir}
    \begin{align}
        (\partial_\tau\xt)^2+(\partial_\s\xt)^2-\left(\frac{\alpha_+}{\xr}\right)^2\left((\partial_\t\xr)^2+(\partial_\s\xr)^2\right)&=0\,,\\
        \partial_\t\xt\partial_\s\xt-\left(\frac{\alpha_+}{\xr}\right)^2\partial_\t\xr\partial_\s\xr&=0\,.
    \end{align}
\end{subequations}
The equations of motion of $\xf$, $\xt$ and $\xr$ are
\begin{subequations}
    \begin{align}
        0=&~\partial_\t^2\left(y^\phi-\frac{J}{2r_+^2}y^t\right)-\partial_\s^2\left(y^\phi-\frac{J}{2r_+^2}y^t\right)+\frac{J}{r_+^3}\left(\partial_\t(\xr\partial_\t\xt)-\partial_\s(\xr\partial_\s\xt)\right)\,,\label{eq:NRC NLO EOM1}\\
        0=&~\partial_\t\left(\frac{4r_+\xr}{\ell^2}\partial_\t\xt\right)+\partial_\s\left(\frac{4r_+\xr}{\ell^2}\partial_\s\xt\right)+J\partial_\t^2\left(y^\phi-\frac{J}{2r_+^2}y^t\right)\nonumber\\&-J\partial_\s^2\left(y^\phi-\frac{J}{2r_+^2}y^t\right)\,,\label{eq:NRC NLO EOM2}\\
        0=&~\partial_\t \left( \frac{\partial_\t \xr }{\xr} \right) - \partial_\s \left( \frac{\partial_\s \xr }{\xr} \right) + \frac{1}{2\a_+^2}\left(\left(\partial_\t \xt \right)^2 - \left(\partial_\s \xt \right)^2 \right)\nonumber\\&+\frac{1}{2 \left( \xr \right)^2 } \left( \left(\partial_\t \xr \right)^2 - \left(\partial_\s \xr \right)^2 \right).\label{eq:NRC NLO EOM3}
    \end{align}
\end{subequations}
At this point, three remarks are in order. To begin with, equation \eqref{eq:NRC NLO EOM3} matches equation \eqref{nlo eom 3} under the substitution $\alpha \to \alpha_+$. Secondly, equations \eqref{eq:NRC NLO EOM1} and \eqref{eq:NRC NLO EOM2} can be linearly combined to obtain
\begin{subequations}
    \begin{align}
        \partial_\t^2\left(y^\phi-\frac{J}{2r_+^2}y^t\right)-\partial_\s^2\left(y^\phi-\frac{J}{2r_+^2}y^t\right)&=0\,,\label{eq:NRC NLO EOM4}\\
        \partial_\t(\xr\partial_\t\xt)-\partial_\s(\xr\partial_\s\xt)&=0\label{eq:NRC NLO EOM5}\,.
    \end{align}
\end{subequations}
Such that equation \eqref{eq:NRC NLO EOM4} is reminiscent of \eqref{nlo eom 1}, and \eqref{eq:NRC NLO EOM5} exactly matches \eqref{nlo eom 2}. Lastly, the NLO Virasoro constraints \eqref{eq:NRC NLO Vir} match the NLO Virasoro constraints obtained in the static case \eqref{nlo vir con} under the substitution $\alpha \to \alpha_+$. These remarks imply that the solutions for $\xt$ and $\xr$ are the same as in the static case, summarised in table \ref{tab:solutions}. The $\xt$ and $\xr$ embedding fields retain the same physical interpretation as in the static case, but the LO transverse embedding field $\xf$ shows a complicated evolution due to the contribution from $\xt$. The NLO embedding fields mix due to the rotation, but still exhibit stringy modes.
\medskip

\begin{figure}[!htb]
   \begin{minipage}{0.48\textwidth}
     \centering
     \includegraphics[width=.95\linewidth]{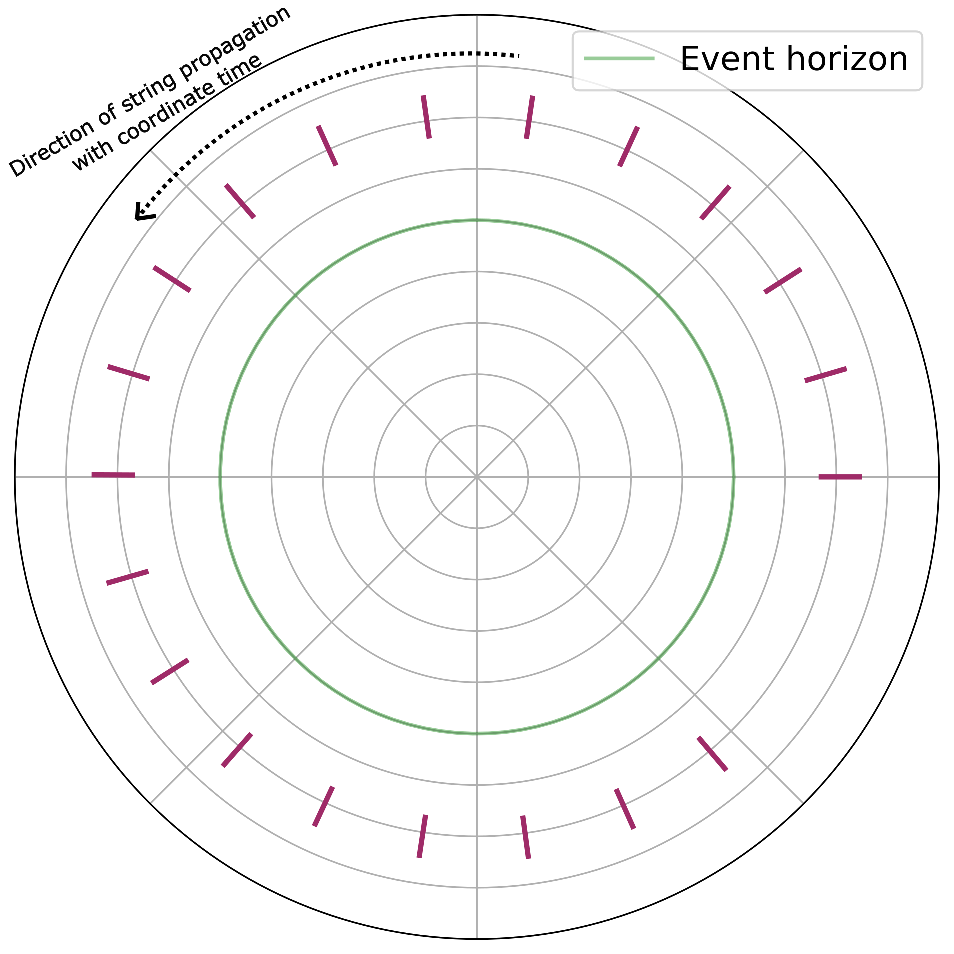}
     \caption{Snapshots of folded string of constant length revolving around the black hole at a fixed radius. The corresponding solution in the static case is given by ansatz \eqref{ansatz1} with $\vartheta\neq0$, i.e. a static yo-yo like solution.}\label{Fig:Data1}
   \end{minipage}\hfill
   \begin{minipage}{0.48\textwidth}
     \centering
     \includegraphics[width=.95\linewidth]{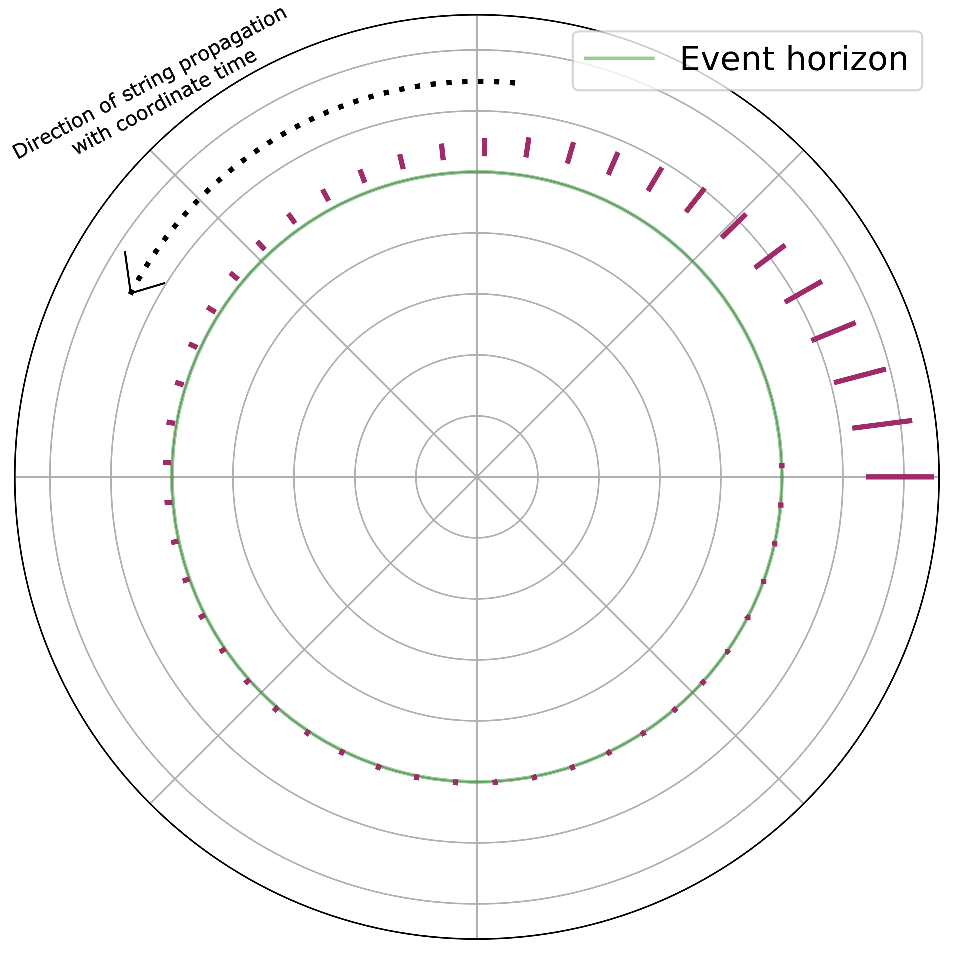}
     \caption{Snapshots of a folded string of decreasing length spiralling towards the black hole. The time-reversed version indicates an elongating string spiralling away from the black hole. The corresponding solution in the static case is given via \eqref{ansatz2} with $\theta\not\in\{0,1\}$ and $\omega\neq0$. }\label{Fig:Data2}
    
   \end{minipage} 
\end{figure}

\begin{figure}[hbt!]
    \centering
    \includegraphics[width=0.46\linewidth]{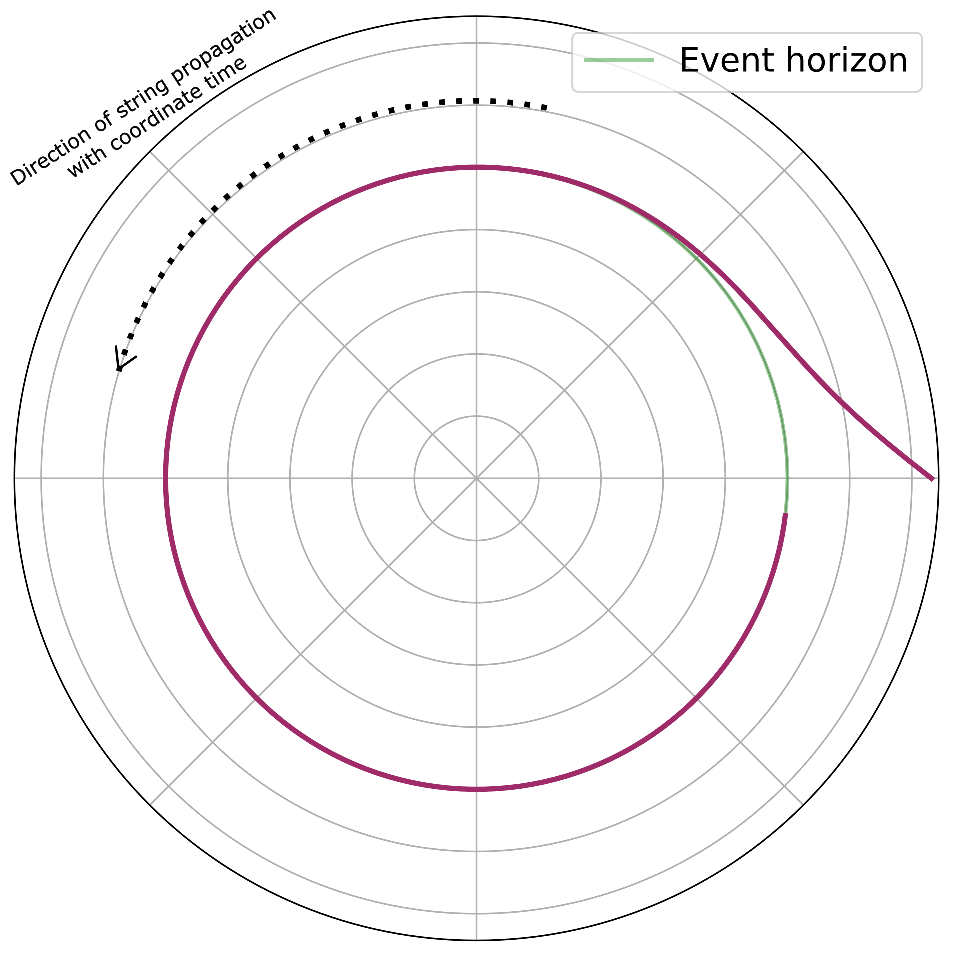}
    \caption{The string becomes a point-like particle and follows a null Rindler geodesic. Its trajectory is described in this figure. The corresponding solution in the static case is given by ansatz \eqref{ansatz1} with $\vartheta=0$.}
    \label{Fig:Data3}
\end{figure}

The diagrams below depict the LO behaviour of strings in the vicinity of a rotating BTZ black hole. The first diagram \ref{Fig:Data1} depicts a folded string of constant length revolving around the black hole at a fixed radius. Whereas, the second diagram \ref{Fig:Data2} shows a folded string of decreasing length spiralling towards the black hole. The time-reversed version of figure \ref{Fig:Data2} indicates the motion of a folded string of increasing length spiralling away from the black hole. Finally, figure \ref{Fig:Data3} shows the string becoming a point-like particle that follows the null Rindler geodesic.     

\subsection{Phase space analysis for magnetic string}\label{Phase space analysis for magnetic string}
The phase space action in string theory $S=\int d\sigma^0L$ describes the characteristics of the string given a background geometry $g_{\mu\nu}$. The phase space Lagrangian $L$ for a relativistic target space is given as \cite{Bagchi:2024rje}
\begin{eqnarray}\label{eq:relativistic phase space action}
    L=\oint d\sigma^1\Big[\dot{X}^\mu P_\mu-\frac{1}{2}e\left(c^2 g^{\mu\nu}(X)P_\mu P_\nu+(c^2T)^2g_{\mu\nu}(X)X'^\mu X'^\nu\right)-uX'^\mu P_\mu\Big],
\end{eqnarray} 
where $e$ and $u$ are Lagrange multipliers. Here, $X'$ and $\dot X$ denotes differentiation with respect to $\sigma^1$ and $\sigma^0$ respectively. 
We will expand $P_\mu$ and $u$ in orders of $c^2$, in addition to the expansion of $X^\mu$ given in \eqref{X expansion}, as follows
\begin{eqnarray}\label{P expansion}
    P_\mu=P_{(0)\mu}+c^2P_{(2)\mu}+\O(c^4),\quad u=u_{(0)}+c^2u_{(2)}+\O(c^4).
\end{eqnarray}
The magnetic Carroll string can be obtained by starting with the scaling $\tilde e:=c^2e$ and $\tilde T:=cT$. Now, the Lagrange multiplier $\tilde e$ is expanded in powers of $c^2$ according to 
\begin{eqnarray}\label{eq:e expansion}
    \tilde e:=\tilde e_{(0)}+c^2\tilde e_{(2)}+c^4\tilde e_{(4)}+\O(c^6).
\end{eqnarray}

\subsection*{LO magnetic phase space theory}
Now that we have all the expansions in place, we can write the LO Lagrangian
\begin{equation}
    \tilde L_{LO} = -\frac{1}{2}\oint  \tilde e_{(0)}\upsilon^{\mu\nu} P_{(0)\mu}P_{(0)\nu}\, d\sigma^1.
\end{equation}
Integrating out $P_{(0)\mu}$ yields $\tilde L_{LO}=0$, rendering the LO theory trivial and devoid of physical content.

\subsection*{NLO magnetic phase space theory}
Moving on to the NLO magnetic string Carroll theory, we find that the action for strings near a static BTZ black hole, after integrating out $P_{(0)\mu}$, is given as, 

\begin{equation}\label{eq:Carroll-BTZ-NLO}
        \tilde S_{\text{NLO}} = \frac{1}{2}\int \oint  \frac{r_h^2}{\tilde e_{(0)}}\left[(\dot x^\phi)^2 + (u_{(0)}^2 - \tilde e^2_{(0)}\tilde T^2 ) (x'^\phi)^2 - 2u_{(0)}\dot x^\phi x'^\phi \right]\,\, d^2\sigma\,.
\end{equation}
The equation of motion for $\tilde e_{(0)}$ yields NLO Virasoro constraints given by,
\begin{eqnarray}
   (\xd^\phi - u_{(0)}x'^\phi)^2 + \tilde e_{(0)}^2 \tilde T^2 (x'^\phi)^2=0.
\end{eqnarray}
Fixing the gauge $u_{(0)} = 0$ and $\tilde e_{(0)} = \text{constant}$ such that $\tilde e_{(0)} \tilde T = 1$ we get,
\begin{equation} \label{EOM:e0}
   (\xd^\phi)^2 + (x'^\phi)^2 = 0 \implies x^\phi=x^\phi_0~~\text{(constant)}.
\end{equation}
The solution \eqref{EOM:e0} also solves the equations of motion associated with $\xf$ and $u_{(0)}$. 

\subsection*{NNLO magnetic phase space theory}
Following the same procedure as in the previous section, we find that the following action governs the NNLO theory,

\begin{equation}
    \begin{split}
        \tilde S_{\text{NNLO}} &= \int \oint\,\frac{1}{2\tilde e_{(0)}}\bigg[ -\left(\frac{2r_h x^\rho}{\ell^2}\right)\big((\dot x^t)^2 - 2u_{(0)}\dot x^t x'^t + (u_{(0)}^2 - \tilde e_{(0)}^2\tilde T^2) (x'^t)^2 \big)\\
        &\quad + \left(\frac{\ell^2}{2r_h x^\rho}\right)\big((\dot x^\rho)^2 - 2u_{(0)}\dot x^\rho x'^\rho + (u_{(0)}^2 - \tilde e_{(0)}^2\tilde T^2) (x'^\rho)^2 \big) \\
        &\quad + 2r_h x^\rho\big((\dot x^\phi)^2 - 2u_{(0)}\dot x^\phi x'^\phi + (u_{(0)}^2 - \tilde e_{(0)}^2\tilde T^2) (x'^\phi)^2 \big)\\
        &\quad + 2r_h^2\big(\dot x^\phi \dot y^\phi - 2u_{(0)}\dot x^\phi y'^\phi + (u_{(0)}^2 - \tilde e_{(0)}^2\tilde T^2) x'^\phi y'^\phi\big)\\
        &\quad + r_h^2\frac{\tilde e_{(2)}}{\tilde e_{(0)}} \big(-(\dot x^\phi)^2 + 2u_{(0)}\dot x^\phi x'^\phi + (-u_{(0)}^2 - \tilde e_{(0)}^2\tilde T^2) (x'^\phi)^2\big)\\
        &\quad + 2 h_{\mu\nu}u_{(2)}\big( -(\dot x^\phi)^2 + u_{(0)} (x'^\phi)^2 \big)\bigg]\,d^2\sigma\,.
    \end{split}
\end{equation}

Varying the NNLO action with respect to $u_{(0)}$ and $\tilde e_{(0)}$, and subsequently imposing our gauge choice, yields the NNLO Virasoro constraints
\begin{subequations}\label{eq:NNLO Virasoro constraints}
\begin{align}
\xd^t x'^t - \left(\frac{\ell^2}{2r_hx^\rho}\right)^2 \xd^\rho x'^\rho &= 0\,,\\
(\xd^t)^2 + (x'^t)^2 - \left(\frac{\ell^2}{2r_hx^\rho}\right)^2 \left[(\xd^\rho)^2 + (x'^\rho)^2\right] &= 0 \,.
\end{align}
\end{subequations}
The equations of motion for $x^\phi$, $x^t$ and $x^\rho$ in our gauge are
\begin{subequations}\label{eq:NNLO EOM}
\begin{align}
\ddot y^\phi - y''^\phi&=0\,,\\
x^\rho (\partial_\tau^2 - \partial_\sigma^2)x^t + (\xd^t \xd^\rho - x'^t x'^\rho) &= 0\,,\\
\partial_\t \left( \frac{\xd^\rho }{\xr} \right) - \partial_\s \left( \frac{x'^\rho }{\xr} \right) + \frac{1}{2}\left(\frac{2r_h}{\ell^2}\right)^2\left[ \left(\xd^t \right)^2 - \left(x'^t \right)^2 \right] \nn \\+\frac{1}{2 \left( \xr \right)^2 } \left[ \left( \xd^\rho \right)^2 - \left(x'^\rho \right)^2 \right]&=0\,.
\end{align}
\end{subequations}
The NNLO equations of motion \eqref{eq:NNLO EOM} and the Virasoro constraints \eqref{eq:NNLO Virasoro constraints} exactly match with the NLO equations of motion \eqref{nlo eom} and Virasoro constraints \eqref{nlo vir con}, which were obtained while analysing NLO Polyakov action. The analysis shown in this section reinforces the claim that, for magnetic strings near a BTZ black hole, the LO and NLO Polyakov action describe the same dynamics as the NLO and NNLO phase space action \cite{Bagchi:2024rje} respectively.  

\section{Strings near BTZ black holes: Electric Theory}\label{BTZ black hole: Electric Strings}
In this section, we will study the dynamics of electric strings in the vicinity of a BTZ black hole. Like the magnetic sector, we will analyse the dynamics of electric strings in two backgrounds, static and rotating (non-extremal). In contrast to the magnetic sector, where both the Polyakov and phase space actions can serve as valid starting points, the electric sector necessitates the use of the phase space action for its analysis.
\medskip

The electric Carroll string can be obtained from relativistic phase space action \eqref{eq:relativistic phase space action} by scaling the Lagrange multiplier and the tension in a way different than the magnetic case: $\hat e:=e$ and $\hat T:=c^2~T$, which are kept fixed through the limit. Playing the same game as before, we will substitute the expansions \eqref{P expansion} and \eqref{eq:e expansion} (now for $\hat e$) along with the above-mentioned scalings into the relativistic phase space action \eqref{eq:relativistic phase space action}. The phase space Lagrangian arranges in orders of $c^2$ and at the LO we get,
\begin{equation}\label{L:gen_electric-LO}
    \hat L_{\text{LO}} = \oint\bigg[ \dot x^\mu P_{(0)\mu} - \frac{1}{2}\hat e_{(0)}\left(\upsilon^{\mu\nu} P_{(0)\mu} P_{(0)\nu} + \hat T^2 \Omega_{\mu\nu} x'^\mu x'^\nu \right) - u_{(0)} x'^\mu P_{(0)\mu} \bigg]\,d\sigma^1.
\end{equation}
It is convenient to decompose the momentum $P_\mu$ into its components along the longitudinal and transverse directions as follows
\begin{equation}
    P_A := v^\mu_A P_\mu\,,\qquad P_{\bar{A}} := e^\mu_{\bar{A}} P_\mu\,.
\end{equation}
When expressed in terms of these projected components, the electric Lagrangian~\eqref{L:gen_electric-LO} depends on $P_{\bar{A}}$ linearly and has a quadratic dependence on $P_A$. This structure permits one to identify $P_{\bar{A}}$ as Lagrange multipliers and integrate out $P_A$ to give the following Lagrangian at the LO\footnote{We have dropped the ``$(0)$" from the subscripts since we will only be analysing the LO electric theory.} 
\begin{equation}\label{L:LO_electriclimit2}
\hat L_{\text{LO}} = \oint \,\bigg[ \left(\dot x^\mu - u x'^\mu\right)e^{\bar{A}}_\mu P_{\bar{A}} + \frac{1}{2\hat e}\tau_{\mu\nu}\left(\dot x^\mu - u x'^\mu\right)\left(\dot x^\nu - u x'^\nu\right) - \frac{\hat T^2}{2}\hat e\, \Omega_{\mu\nu} x'^\mu x'^\nu \bigg] d\sigma^1\,.
\end{equation}

The worldsheet gauge redundancy can be fixed globally by setting \cite{Polchinski:1998rq,Bagchi:2024rje}
\begin{eqnarray}\label{eq: electric strings gauge}
\hat e=1,\quad u(\sigma^0,\sigma^1)=u(\sigma^0).
\end{eqnarray}
Though locally we can fix $\hat e=1$ and $u=0$, i.e. the conformal gauge, but globally this becomes a problem, which is true even for relativistic strings. Since we are dealing with closed strings, the worldsheet is a cylinder parameterised by $(\sigma^0,\sigma^1)$. Where $\sigma^1$ is compact in the domain $[0,1)$ and $-\infty<\sigma^0<\infty$ in any coordinate system with which we parameterise the cylinder. Note that a worldsheet diffeomorphism has the form $\delta\sigma^a=-\zeta^a$, after which, to keep $0\leq\sigma^1<1$, $\zeta^1$ should have a periodicity of $1$, and the contribution of its zero mode should vanish to keep the origin fixed.
However, diffeomorphisms of the form $\sigma^1\to\sigma^1+f(\sigma^0)$ are available to us. But we will not use it to fix the gauge to $u=0$ because we do not want the limits of the integral to depend on an arbitrary function of $\sigma^0$. Rather, we want it to be fixed between $0$ and $1$. Therefore, for a globally defined worldsheet, we would have to compromise to a partial gauge fixing like in \eqref{eq: electric strings gauge}, something we need to be careful of in a generic electric solution. In what follows, however, we would discuss the residual gauge symmetries on the electric worldsheet, but by explicitly using the conformal gauge. 
\medskip

To make this argument more precise, the passage from the global gauge~\eqref{eq: electric strings gauge} to the local conformal gauge can be implemented by the following reparametrisation:
\begin{equation}\label{eq: electric string local gauge}
    \tilde{\sigma}^0 = \sigma^0, 
    \qquad 
    \tilde{\sigma}^1 = \sigma^1 + U(\sigma^0)\,,
\end{equation}
where the function \( U(\sigma^0) \) is related to the original Lagrange multiplier via
\begin{equation}
    \tilde{\partial}_{0} U(\tilde{\sigma}^0) = u(\tilde{\sigma}^0)\,.
\end{equation}
This reparametrisation renders the subsequent equations of motion considerably more tractable, as will become evident in \eqref{section: static BTZ electric}. So, the takeaway is, we will be zooming in on the local physics on the electric worldsheet without loss of any generality in what follows.

\subsection{Residual symmetries on the worldsheet}\label{sec:residual symmetries}
Now that we have discussed the structure of the LO electric action and gauge fixing, we should ask what symmetry arises as the residual symmetry group after gauge fixing in this case? Note that the magnetic string had a relativistic worldsheet, which would not be true in this case. The diffeomorphism invariance of the electric action becomes manifest upon formulating it in terms of the zweibeine $(\mathbb{p}_a,\mathbb{h}_a)$ and its inverse $(\mathbb{q}^a,\mathbb{h}^a)$, defined as
\begin{eqnarray}\label{zweibeine}
    u = - \frac{\mathbb{q}^1}{\mathbb{q}^0},~~~~\hat e = \frac{1}{(\mathbb{q}^0)^2\mathbb{h}}\,,~~~~\mathbb{h}=\det(\mathbb p_a,\mathbb h_a)\,.
\end{eqnarray}
The zweibeine and its inverse in two dimensions are related as follows
\begin{subequations}
\begin{gather}
    \mathbb{q}^0= -\frac{\mathbb{h}_1}{\mathbb{h}},~~~\mathbb{q}^1= \frac{\mathbb{p}_0}{\mathbb{h}},~~~\mathbb{h}^0 = -\frac{\mathbb{p}_1}{\mathbb{h}},~~~\mathbb{h}^1 = \frac{\mathbb{p}_0}{\mathbb{h}},~~~ \mathbb{h} = \mathbb{p}_0 \mathbb{h}_1 - \mathbb{p}_1 \mathbb{h}_0\,.
\end{gather}
\end{subequations}
Using the zweibeine formulation in \eqref{zweibeine}, the Lagrangian \eqref{L:LO_electriclimit2} can be written more compactly:
\begin{equation}\label{L:general}
    \hat L_{LO}
= \oint \mathbb{h}
\Big[
\hat T^2\,\mathbb{q}^a \partial_a x^\mu\,e_\mu^{\bar{A}}\,\tilde P_{\bar{A}}
+\frac{1}{2}\,\tau_{\mu\nu}\,\mathbb{q}^a\partial_a x^\mu\,\mathbb{q}^b\partial_b x^\nu
-\frac{\hat T^2}{2}\,\Omega_{\mu\nu}\,\mathbb{h}^a\partial_a x^\mu\,\mathbb{h}^b\partial_b x^\nu
\Big]\, d\sigma^1,
\end{equation}
with $\tilde P_{\bar{A}}$ defined to be $\tilde P_{\bar{A}} = -\cfrac{1}{\hat T^2 \mathbb h_1}P_{\bar{A}}$. Analogue of this action was already found in \cite{Bagchi:2023cfp}, and it was promptly noted that setting $\hat T=0$, we get the action for null strings \cite{Bagchi:2015nca,Isberg:1993av,Bagchi:2013bga}. Therefore, the action \eqref{L:general} can be thought of as tensile deformations of the tensionless action. Observe that in \eqref{L:LO_electriclimit2}, the third term has no $\sigma^0$ derivatives of the LO embedding fields. Hence, we need to gauge fix $\mathbb{h}^a = (0,\mathbb{h}^1)$ in \eqref{L:general} so as to recover the LO electric Lagrangian as written in \eqref{L:LO_electriclimit2}. This follows due to the local Carroll boost invariance of the worldsheet, which acts on the zweibeine and $\tilde P_{\bar{A}}$ as 
\begin{eqnarray}
\delta\mathbb p_a=\lambda\mathbb h_a\,,~~~\delta\mathbb h_a=0\,,~~~\delta \mathbb q^a=0\,,~~~\delta \mathbb h^a=\lambda\mathbb q^a\,,~~~\delta\tilde P_{\bar{A}}=\lambda\mathbb{h}^a\partial_ax^\mu e_{\mu\bar A}\,,
\end{eqnarray}
where $\lambda(\s)$ is a function of the worldsheet coordinates. The geometry on the worldsheet is naturally reparametrisation invariant. The diffeomorphisms are generated by $\zeta^a$ and act on the fields as
\begin{eqnarray}
    \delta\mathbb p_a=\mathcal{L}_\zeta\mathbb p_a\,,~~~\delta\mathbb h_a=\mathcal{L}_\zeta\mathbb h_a\,,~~~\delta x^\mu=\zeta^a\partial_a x^\mu\,,~~~\delta \tilde P_{\bar{A}}=\zeta^b\partial_b\tilde P_{\bar{A}}\,,
\end{eqnarray}
similarly for the inverse zweibeine. Finally, the worldsheet is invariant under Weyl rescaling as well, which acts on the zweibeine and $\tilde P_{\bar{A}}$ as
\begin{eqnarray}
\delta\mathbb p_a=\omega\mathbb p_a\,,~~~\delta\mathbb h_a=\omega\mathbb h_a\,,~~~\delta \mathbb q^a=-\omega\mathbb q^a\,,~~~\delta \mathbb h^a=-\omega\mathbb h^a\,,~~~\delta\tilde P_{\bar{A}}=-\omega \tilde P_{\bar{A}}.
\end{eqnarray}
The couplings in the Lagrangian \eqref{L:general} transform under the target space string-Carroll boost transformation with $\lambda^A_{(0)\bar B}$ (cf. \cite{Bagchi:2024rje}) being the transformation parameter
\begin{eqnarray}
    \delta\tilde P_{\bar{A}}=-\eta_{AB}\mathbb{q}^a\partial_ax^\mu\tau_\mu^B\lambda^A_{(0)\bar A}\,,~~~\delta\tau^A_\mu=\lambda^A_{(0)\bar B}e^{\bar B}_\mu\,.
\end{eqnarray}
Under these transformations, the Lagrangian for the electric Carroll strings \eqref{L:general} can be shown to transform as a total derivative, which proves that the worldsheet theory is invariant under local Carroll boosts, Weyl rescalings and diffeomorphisms.
\medskip

The gauge condition $\mathbb h^a = (0,\mathbb h^1)$ implies that $\delta\mathbb h^0=0$. Using the transformation properties of $\mathbb h^a$ a relation between $\lambda$ and $\zeta^a$ can be deduced such that
 \begin{equation}
     \lambda = \frac{\mathbb h^1}{\mathbb q^0} \partial_1 \zeta^0.
 \end{equation}
Furthermore, the transformation laws of $\hat e$ and $u$ can be deduced using the relations \eqref{zweibeine}
\begin{subequations}
    \begin{align}
    \delta u & = \zeta^a \partial_a u - u^2 \partial_1 \zeta^0 + u \left( \partial_0\zeta^0 - \partial_1 \zeta^1 \right) + \partial_0\zeta^1, \\
    \delta \hat e & = \zeta^a \partial_a \hat e - 2\hat e u\partial_1\zeta^0 + \hat e \left( \partial_0\zeta^0 - \partial_1 \zeta^1 \right).
    \end{align}
\end{subequations}
Lastly, as discussed earlier, the gauge redundancy can be fixed locally by setting $\hat e=1$ and $u=0$. The transformations that preserve this gauge are determined by requiring that the gauge-fixing conditions remain invariant, that is, $\delta \hat e = \delta u=0$. Solving these conditions, we get\footnote{Here, we have relabelled the coordinates as $\sigma^0\to\tau$ and $\sigma^1\to\sigma$ for a more familiar structure.},
\begin{subequations}\label{residual diffeomorphism}
\begin{align}
    \zeta^a & = \left( f'(\s)\t + g(\s), f(\s) \right)\, ,\\
    \lambda & = \left( \frac{\mathbb h^1}{\mathbb q^0} \right) \left( f''(\s)\t + g'(\s) \right),
\end{align}
 \end{subequations}
where $f,g$ are arbitrary functions depending on $\s$. The action of $\zeta^a$ in an arbitrary function $\Psi(\tau,\sigma)$ is given as,
\begin{eqnarray}
\delta\Psi=\Big[f'(\s)\t\partial_\t+f(\s)\partial_\s\Big]\Psi+g(\s)\partial_\t\Psi=\Big[L(f)+M(g)\Big]\Psi\,,
\end{eqnarray}
where, $L(f)$ and $M(g)$ are defined as the vector fields
\begin{eqnarray}
    L(f) = f'(\sigma)\tau\partial_\tau + f(\sigma)\partial_\sigma\,,\,\,\,\, M(g) = g(\sigma)\partial_\tau.
\end{eqnarray}
The generators $L(f)$ and $M(g)$ are found to satisfy the following algebra
\begin{align}
    \left[L(f_1), L(f_2)\right] = &L(f_1f_2'-f_1'f_2),\quad
    \left[L(f), M(g)\right] = M(fg'-f'g), \nonumber\\
    &~~~~~~\left[M(g_1), M(g_2)\right] = 0.
\end{align}
We decompose the functions $f,g$ in terms of Fourier modes as
\begin{equation}
    f(\sigma) = \sum_n c_ne^{in\sigma}, \qquad g(\sigma) = \sum_n d_ne^{in\sigma},
\end{equation} can be written in a much suggestive form:
\begin{align}\label{BMS3 Algebra}
    \left[L_m, L_n\right] = (m-n)L_{m+n}\,,~~~
    \left[L_m, M_n\right] = (m-n)M_{m+n}\,,~~~
    \left[M_m, M_n\right] = 0,
\end{align}
Where we have used mode expansions of the generators $L(f)$ and $M(g)$ on the cylinder
\begin{subequations}
    \begin{align}
        L(f) &= \sum_n c_n e^{in\sigma}\left(\partial_\sigma + in\tau\partial_\tau\right)=-i\sum_n c_nL_n\,,\\
        M(g) &= \sum_n d_n e^{in\sigma}\partial_\tau=-i\sum_n d_n M_n\,.
    \end{align}
\end{subequations}
The algebra derived in equation \eqref{BMS3 Algebra} can be identified with the $\text{BMS}_3$, or equivalently, the two-dimensional conformal Carroll algebra with $z=1$ \cite{Duval:2014uva,Hao:2021urq,Bagchi:2015nca,Figueroa-OFarrill:2025njv}, sans the central terms. Therefore, the worldsheet theory is invariant under the transformations generated by the $\text{BMS}_3$ group.
\medskip

Before moving on, we should focus on this result a bit.
Note that, as shown in literature, the residual gauge symmetry group on the worldsheet for null strings also turns out to be the $\text{BMS}_3$ group \cite{Isberg:1993av, Bagchi:2013bga}. The null strings, however, are obtained from the ILST action \cite{Isberg:1993av}, which occurs as a tensionless limit of the tensile string action. In contrast, the Lagrangian \eqref{L:general}, in our case, contains a tensile deformation along with the tensionless part, and only reduces to the null string when $\hat{T}=0$. The tensile deformation terms at order $\hat{T}^2$, as one can notice, manifests as a magnetic scalar theory in the transverse direction. So from the worldsheet perspective, it is imperative to postulate that the extra terms act as a \textit{Carroll marginal} deformation (after gauge fixing) to our theory, generalising the class of null strings beyond pure tensionless strings. The physical picture, especially the spectrum associated to such deformations clearly need additional investigation.     

\subsection{Electric solutions near BTZ black hole} \label{section: static BTZ electric}
In this subsection, we will go back to our action and analyse classical electric string solutions near a static BTZ black hole. Using our string Carroll expanded fields, the explicit form of the LO phase space Lagrangian is given as 
\begin{multline}\label{lagrangian:BTZ}
    \hat{L}_{LO} = \oint \Bigg[\frac{1}{2\hat{e}} \left\{ -\frac{x^\rho}{\alpha}\left(\xd^t - u x'^t\right)^2 + \frac{\alpha}{x^\rho}\left(\xd^\rho - u x'^\rho\right)^2\right\}\\+r_h (\xd^\phi - ux'^\phi)P-\frac{\hat{T}^2}{2}\hat{e} r_h^2 (x'^\phi)^2 \Bigg] \, d\sigma^1.
\end{multline}
The equations of motion for $P$, $u$, $\hat{e}$, $x^\phi$, $x^t$, $x^\rho$ are\footnote{Note that the tension terms appear here as source terms in Hamiltonian constraint and the same for angular evolution.}
\begin{subequations}\label{eq: LO Electric strings EOM}
\begin{align}
\xd^\phi - ux'^\phi&=0\label{EOM P}\\
-r_h x'^\phi P + \frac{1}{\hat{e}}\left[\frac{x^\rho}{\alpha}x'^t (\xd^t - ux'^t) - \frac{\alpha}{x^\rho}x'^\rho (\xd^\rho - ux'^\rho)\right]&=0\label{EOM u}\\
\frac{1}{2\hat{e}^2}\left[-\frac{x^\rho}{\alpha}\left(\xd^t - ux'^t\right)^2 + \frac{\alpha}{x^\rho} \left(\xd^\rho - ux'^\rho\right)^2\right] + \frac{\hat{T}^2 r_h^2}{2} (x'^\phi)^2 &=0\label{EOM e}\\
r_h \dot{P} - r_h (uP)' - \hat{T}^2 r_h^2 (\hat{e}x'^\phi)' &=0\label{EOM xf}\\
-\partial_0 \left(\frac{x^\rho}{\alpha\hat{e}}(\xd^t - ux'^t)\right) + \left(\frac{ux^\rho}{\alpha\hat{e}}(\xd^t - ux'^t)\right)'&=0\label{EOM xt}\\
-\frac{1}{2\hat{e}\alpha} (\xd^t - ux'^t)^2 - \frac{\alpha}{2\hat{e} (x^\rho)^2} (\xd^\rho - ux'^\rho)^2- \partial_0 \left(\frac{x^\rho}{\hat{e}\alpha}(\xd^\rho - ux'^\rho)\right) \nonumber\\ + \left(\frac{ux^\rho}{\alpha\hat{e}} (\xd^\rho - ux'^\rho)\right)'&=0\label{EOM xr}
\end{align}    
\end{subequations}
As already stated, it is convenient to solve the equations of motion \eqref{eq: LO Electric strings EOM} in the tilde worldsheet coordinates (defined in \eqref{eq: electric string local gauge}).

The solution for \eqref{EOM P} gives the angular profile
\begin{eqnarray}\label{eq: Electric String xf solution}
    \xf=\xf(\tilde \sigma^1).
\end{eqnarray}
The equation \eqref{EOM xt} can be integrated:
\begin{equation}\label{eq: Electric String xt solution}
    \dot{x}^t = \frac{f(\tilde \sigma^1)}{x^\rho}\, .
\end{equation}
Substituting \eqref{eq: Electric String xt solution} into \eqref{EOM e} and \eqref{EOM xr} yields
\begin{align}
    \frac{1}{\alpha x^\rho} \left( (\alpha\dot{x}^\rho)^2 - f^2 \right) + \hat{T}^2 r_h^2 \left( x'^\phi \right)^2&=0, \label{eqn:eom3_v1} \\
    2x^\rho \ddot{x}^\rho - (\dot{x}^\rho)^2 + \frac{f^2}{\alpha^2}&=0. \label{eqn:eom6_v1}
\end{align}
The equation \eqref{eqn:eom6_v1} can be analysed using the following radial solution
\begin{equation}\label{ansatz}
    x^\rho(\tilde\sigma^0, \tilde\sigma^1) = -(\tilde\sigma^0)^2\frac{A^2}{4\alpha} + \tilde\sigma^0B + C\,,
\end{equation}
where $A$, $B$ and $C$ are functions of $\tilde\sigma^1$ and satisfy the relation 
\begin{equation}
    \alpha^2 B^2+\alpha A^2C = f^2\,.
\end{equation}
Furthermore, substituting the ansatz for $x^\rho$ \eqref{ansatz} into \eqref{eqn:eom3_v1} we get, 
\begin{equation}\label{eq:phi_prime}
    \left(x'^\phi\right)^2 = \frac{A^2}{\hat{T}^2 r_h^2}.
\end{equation}
Therefore, the solutions to \eqref{eq: LO Electric strings EOM} are given as 

\begin{subequations}\label{eq:solutions to worldsheet embedding coordinates}
\begin{align}
    \xt (\tilde \sigma^0,\tilde \sigma^1) &=\alpha\log\left|\frac{\tilde\sigma^0 A(\tilde \sigma^1)^2-2(\alpha B(\tilde \sigma^1) - f(\tilde \sigma^1) )}{\tilde\sigma^0 A(\tilde \sigma^1)^2 - 2(\alpha B(\tilde \sigma^1) + f(\tilde \sigma^1) )}\right| + E(\tilde \sigma^1)\,,  \label{eq:solutions to worldsheet embedding coordinates 1}\\
    %\frac{ f }{A } \ln \left| \frac{ [ A^2 \tilde \sigma^0 - 2\a B ] - f }{ [ A^2 \tilde \sigma^0 - 2\a B ] + f } \right| + E(\tilde \sigma^1)\,,\\
	\xr(\tilde \sigma^0,\tilde \sigma^1) &= - \frac{A(\tilde \sigma^1)^2}{4\a} \left(\tilde \sigma^0\right)^2 + B(\tilde \sigma^1) \tilde \sigma^0 + C(\tilde \sigma^1)\,, \label{eq:solutions to worldsheet embedding coordinates 2}\\
	\xf(\tilde \sigma^0,\tilde \sigma^1) &= \frac{\int A(\tilde \sigma^1)\, d\tilde \sigma^1}{r_h \hat{T} }\,, \label{eq:solutions to worldsheet embedding coordinates 3}\\
	%\label{sol end}
	P(\tilde \sigma^0,\tilde \sigma^1) &= \hat{T}A'(\tilde \sigma^1)\tilde \sigma^0 + D(\tilde \sigma^1)\,. \label{eq:solutions to worldsheet embedding coordinates 4}
\end{align}
\end{subequations}
$A$, $B$, $C$, $D$, $E$ and $f$ are arbitrary functions of $\tilde\sigma^1$, appearing as integration constants\footnote{We can choose these arbitrary functions consistently such that the LO solutions \eqref{eq:solutions to worldsheet embedding coordinates} satisfy the equations of motion \eqref{eq: LO Electric strings EOM}. For some choices of $A$, $B$, $C$, $D$, $E$ and $f$, it can turn out that the LO solutions are not defined on the whole worldsheet, but rather only on a subset of it. This observation reinforces our claim that the worldsheet coordinate transformation \eqref{eq: electric string local gauge} actually transforms a global chart $(\sigma^0, \sigma^1)$ to a local chart $(\tilde \sigma^0, \tilde \sigma^1)$ on the worldsheet.}. The exact dependence of the longitudinal embedding coordinates on $\tilde\sigma^1$ cannot be deduced at the LO. But they do represent the initial data that determines the motion of the string. However, some features of the electric strings can still be inferred from the LO analysis. Setting $\dot P=0$ for simplicity, we find that $\xf=\tilde\sigma^1/(r_hT)$, implying that the string linearly wraps around the black hole and follows the geodesic on $S^1$. This solution implies that the string goes around the horizon, which points to the elongation of extended objects in the near-horizon regime of black holes. Similar studies of strings in the vicinity of black holes and in accelerated frames \cite{deVega:1987um,Susskind:1993aa}, also report the transverse elongation of strings. However, note that the inverse tension term acts as a winding for the $x^\phi$, which clearly shows that small tension strings wound multiple times around the horizon.

\subsubsection*{Rotating BTZ }
In the magnetic section, we have clearly shown that in the co-rotating frame, the near-horizon expansion of the rotating BTZ black hole is exactly like in the static case. The story, and the mapping of solutions remain same in the electric case as well. We obtain the same equations of motion as the static case with $\xt$, $\xf$, $r_h$ and $\a$ replaced with $\xtt$, $\xff$, $r_+$ and $\a_+$ defined by \eqref{non extremal coordinate transform} and \eqref{alpha plus}. Hence, in the co-rotating frame, the string retains the same physical description as in the cases described above. Hence, we refrain from discussing those in details. 

\section{Generic features of $2+1$ D solutions}\label{Features of three and four dimensions}
Now that we have gone through a detailed discussion on our solutions, lets focus on some particular structures associated to our solutions. 
 Owing to the similarity in the near-horizon expansion of the 3+1 dimensional Schwarzschild and 2+1 dimensional static BTZ black holes, one might anticipate a similar string dynamics in their vicinity. However, this expectation holds partially. In both instances, the magnetic string is confined to the longitudinal direction and exhibits constrained dynamics. Nonetheless, we have identified a broader class of solutions than those presented in \cite{Bagchi:2024rje} thanks to the analytical tractability in this case. While \cite{Bagchi:2024rje} show the existence of folded/yo-yo strings, we have shown that these yo-yo strings can either have a constant radial extent or exhibit time-dependent radial evolution. But these folded strings and geodesics exhaust the parameter space in the magnetic case. Of course, the caveat is that we have not considered, say, a non separable ansatz for the solutions, but we should not be missing any physics here. 
\medskip

The story is more interesting in the case of electric strings, which exhibit different behaviour owing to the topological differences between a Schwarzschild and a BTZ black hole.
In the electric sector, we have obtained a solution where the string wraps around the event horizon of the BTZ black hole. This is reminiscent of the electric string solutions explored in \cite{Bagchi:2024rje}, but there are slight differences between the two solutions. Firstly, in the case of the 3+1-dimensional Schwarzschild black hole, as the electric string falls towards the black hole, it can show one of two behaviours.
\begin{figure}[hbt!]
\centering
\includegraphics[width = \linewidth]{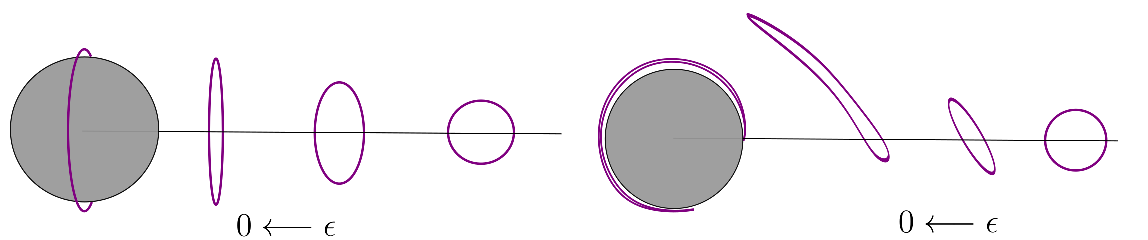}
\caption{\textbf{Left}: An electric string encircling a 3+1 dimensional Schwarzschild black hole. \textbf{Right}: A two-dimensional spatial projection of an electric string wrapping around a 3+1-dimensional Schwarzschild black hole. }
\label{fig: electric strings near Schwarzschild Black hole}
\end{figure}
\begin{figure}[hbt!]
\centering
\includegraphics[width = 0.6\linewidth]{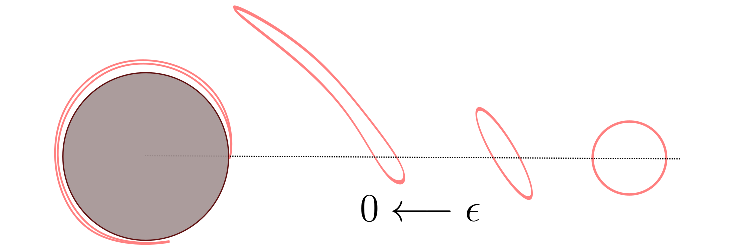}
\caption{An electric string wrapping around a 2+1-dimensional BTZ black hole.}
\label{fig: electric strings near BTZ Black hole}
\end{figure}
The electric string can either encircle the black hole, like a rubber band encircling a ball. The electric string, however, can also wrap around one of the great circles. The wrapping solution demonstrates the elongation of strings in the transverse direction. This solution requires a forcing mechanism which forces the string to keep lying (or wrap multiple times) on one of the great circles, or on a simply connected manifold it is simply allowed to shrink to a point. The former interpretation is depicted on the left in figure \ref{fig: electric strings near Schwarzschild Black hole} and the latter is depicted on the right.
\medskip

 In the case of the 2+1-dimensional BTZ black hole, owing to the topology of $S^1$, the electric string can only wrap around the black hole as depicted in figure \ref{fig: electric strings near BTZ Black hole}. But the reader could ask, what guarantees that the wrapped string will stay wrapped on the black hole? What constrains it to perhaps stay wounded?
\medskip

Motivated by the question above, we now turn to a discussion of electric string solutions and their winding on the horizon of the static BTZ black hole. Note that equations \eqref{eq: Electric String xf solution} and \eqref{eq:solutions to worldsheet embedding coordinates} imply that $\xf$ has no temporal dynamics. Hence, the string's angular configuration is expected to be frozen in coordinate time. The equation of motion of $\xf$ \eqref{EOM xf} in the tilde coordinates \eqref{eq: electric string local gauge} along with the gauge fixing condition \eqref{eq: electric strings gauge} is given as 
\begin{equation}\label{eq: xphi almost geodesic}
    \frac{d}{d\tilde\sigma^1}\lb \frac{d \xf(\tilde\sigma^1)}{d\tilde\sigma^1}\rb = \frac{\tilde\partial_0 P(\tilde\sigma^0,\tilde\sigma^1)}{r_h \hat T^2} \,,
\end{equation}
where it is evident that the RHS works as a source term for the spatial dynamics along $x^\phi$\footnote{In fact, one can look at \eqref{EOM xf} and naively compare this with a continuity equation in 1+1 D, where the angular momentum $P$ is a current density and $-uP$ is convective current associated to that when one thinks of $u$ as a `velocity'. The $\hat{T}^2$ term acts as a source which provides elastic deformations.}. It is noteworthy that the $\tilde\partial_0 P$ term signifies a change in angular momentum and from the discussion around \eqref{eq:solutions to worldsheet embedding coordinates} we could see the (non)-vanishing of this term signified whether there is spatial variation in the winding or not. This prompts us to analyse the wrapping dynamics for these two distinct cases of $\tilde\partial_0 P$ as below.
\medskip

\textbf{\emph{$\blacksquare$ Case $\tilde\partial_0 P= 0$:}} In this case, \eqref{eq: xphi almost geodesic} yields us
\begin{equation}\label{eq: phi affine geodesic}
    \frac{d}{d\tilde\sigma^1}\lb \frac{d \xf(\tilde\sigma^1)}{d\tilde\sigma^1}\rb = 0 ~~~~ \implies \xf(\tilde\sigma^1) = \xf_0 + \omega\tilde\sigma^1\,, ~~\omega \in 2\pi\mathbb Z\,.
\end{equation}
This describes an affinely-parametrised spatial geodesic on $S^1$. This implies that the string wraps on the event horizon of the static BTZ black hole with uniform winding in the angular direction. Since angular momentum density is a constant through the solution, this could be dubbed as an equilibrium configuration where the string behaves like a smooth hoop encircling the horizon.  
\medskip

\textbf{\emph{$\blacksquare$ Case $\tilde\partial_0 P \neq 0$:}}
This is definitely the richer dynamical situation due to the non-uniformity in momentum density.
Consider the following equations: first, \eqref{eq: xphi almost geodesic} with the substitution from \eqref{eq:solutions to worldsheet embedding coordinates 4}; and second, the equation of motion for the $\xf$ field, given in \eqref{eq:phi_prime}
\begin{subequations}\label{eq: P dot not zero}
    \begin{align}
        \partial^2_{\tilde\sigma^1} \xf(\tilde\sigma^1) &= \frac{A'(\tilde\s^1)}{r_h \hat T} \,, \label{eq: P dot not zero 1} \\
        \partial_{\tilde\sigma^1} \xf(\tilde\sigma^1) &= \frac{A(\tilde\s^1)}{r_h \hat T}\,. \label{eq: P dot not zero 2}
    \end{align}
\end{subequations}

If we assume that \eqref{eq: P dot not zero 1} describes a reparametrisation of an affinely-parametrised geodesic, where $\tilde\sigma^1(s)$ is a reparametrisation of an affine parameter $s$, then from \eqref{eq: phi affine geodesic}, $\xf(\sigma)$ will take the form
\begin{equation}
    \xf(\tilde\sigma^1) = \xf_0 + \omega\, s(\tilde\sigma^1) \,, ~~~~\omega \in 2\pi\mathbb{Z}\,.
\end{equation}

From this expression, we get that $\frac{d \xf}{d \tilde\sigma^1}$ does not change sign. We will relabel $\tilde \sigma^0$ as $\tau$ and $\tilde \sigma^1$ as $\sigma$ to think from a worldsheet perspective. This aforementioned condition can then be compactly expressed as
\begin{equation}\label{eq: necessary condition 1}
    \frac{d \xf}{d \sigma} \neq 0\,,~~~ \forall \sigma \,.
\end{equation}

This condition is necessary and sufficient for $\xf(\sigma)$ to be a reparametrisation of an affinely-parametrised geodesic of $S^1$. Note that $\xf(\sigma)$ is solved by \eqref{eq:solutions to worldsheet embedding coordinates 3} and thus condition \eqref{eq: necessary condition 1} translates to
\begin{equation}\label{eq: necessary condition 2}
    A(\sigma) \neq 0 \,, ~~~ \forall \sigma \,.
\end{equation}

We would now like to rewrite this condition from a different perspective. $\xf$ is an angular coordinate with winding and thus it should satisfy $\xf(\s + 1) = \xf(\s) + 2 n\pi $, for some winding number $n \in \mathbb{Z}$. This allows $\xf(\s)$ to have the following mode expansion\footnote{Notice that this mode expansion looks like a naive ultra-relativistic contraction ($\tau\to\epsilon\tau,~\sigma\to\sigma, \epsilon \to 0$) of relativistic bosonic string solutions. Such solutions were considered with respect to near-horizon BTZ black holes in \cite{Grumiller:2019tyl}. }
\begin{equation}
    \xf(\sigma) = \xf_0 + 2n\pi\s + \sum_{m\neq 0}\frac{ \alpha^\phi_m }{ m } e^{2m\pi i\s}\,, ~~~~~~~~~\alpha^\phi_{-m} = -(\alpha^\phi_m)^*\,.
\end{equation}
Note that we are not yet trying to quantise the string, but simply looking at the Fourier modes. Thus, using \eqref{eq: P dot not zero 2}, $A(\s)$ can be written as
\begin{equation}
    A(\s) = 2\pi r_h \hat T \lb n + i\sum_{m\neq 0}\alpha^\phi_m e^{2m\pi i\s}  \rb \,.
\end{equation}

Thus, the necessary and sufficient condition \eqref{eq: necessary condition 2} for $\xf(\sigma)$ to be a reparametrisation of an affinely-parametrised geodesic becomes in terms of $\xf$ modes
\begin{equation}\label{eq: reparametrisation condition}
    i\sum_{m\neq 0}\alpha^\phi_m e^{2m\pi i\s} \neq-n\,, ~~~\forall \sigma\,.
\end{equation}

This condition enforces a monotonic winding, i.e. either $x'^{\phi}(\s)>0~ \forall \s$ or $x'^{\phi}(\s)<0~ \forall \s$, but abhors a change of sign.

\medskip

On the other hand, if equations \eqref{eq: P dot not zero} do not define such a reparametrisation, i.e. \eqref{eq: necessary condition 2} fails, we need to look at the solutions a bit more closely. First, we note that any well-behaved curve on $S^1$ describes a piecewise geodesic on $S^1$. Secondly, in the subset of $[0,1]$, where $A(\sigma)$ does not change sign, \eqref{eq: P dot not zero 1} describes a reparametrisation of an $S^1$ geodesic. Thus, the points of sign flips cause something drastic to the string profile. Since we do not know the exact form of the undetermined functions of $\sigma$ appearing in \eqref{eq:solutions to worldsheet embedding coordinates}, we can hypothesise two possibilities. The first one comes naturally: as the function $A$ changes sign, the string suddenly changes the wrapping direction, which could cause the string to develop kinks at the aforementioned point(s). This would give rise to situations akin to a baklava, where the string arranges itself as layers over various parts of itself. In the other situation, it may so happen that the string might lose its `grip' and leave the $S^1$ wherever the sign changes happen. However, the physical processes associated with these possibilities have to be understood better, and this will appear in a later communication. 

\medskip

However, naively this gives one the idea that our horizon wrapping strings could be of two classes depending on whether they preserve winding or not, giving a notion of topological stability to these solutions. A direct consequence of this structure could mean that different winding solutions cannot really morph into each other. This could remind one of the `horizon strings' construction in \cite{Bagchi:2022iqb}, which also wrap the $\phi$ direction of the BTZ black hole with quantum mechanically separated sectors of non-zero winding, and an infinitely degenerate sector for zero winding. However, we should immediately note these horizon strings are explicitly null ($\hat{T}=0$), which is not our case, so this similarity is a pure coincidence at this point. 

\medskip

For completeness, if one were to compare between $2+1$ and $3+1$ case, they should also note here that Schwarzschild black hole is locally Minkowski, whereas the BTZ black hole is locally AdS. Even though the asymptotic structure is lost on taking the near-horizon limit on the metric, the local features should still be intact. This is reflected in the fact that the near-horizon metrics of the static and rotating BTZ black hole,  \eqref{static expansion 1} and \eqref{non extremal expansion 1}, have a factor $\alpha$ and $\alpha_+$ in the case of the rotating black hole, which has the implicit information of the AdS radius $\ell$.
\medskip

Finally, there are qualitative differences in the string solution near a static versus a rotating BTZ black hole in the non-co-rotating frame \eqref{eq:rotating BTZ black hole map}. In the static case, we see that the LO transverse embedding field freezes on the sphere, whereas in the case of the rotating BTZ black hole, the LO transverse embedding field is dependent on the LO longitudinal embedding field $x^t$. This dependence makes the folded strings and the particle-like strings revolve around the black hole. This feature is due to the frame-dragging effect of the rotating black hole.

\section{Concluding Remarks and Future Directions} \label{Concluding Remarks and Future Directions}
In this work, we started off with the idea that the near-horizon expansion of the three-dimensional non-extremal BTZ black hole naturally takes the form of a string Carroll expansion, with the longitudinal sector forming a two dimensional Rindler spacetime and the transverse sector comprising a compact circle ($S^1$). This shared the intuition obtained for four-dimensional Schwarzschild and Kerr black hole in earlier works \cite{Bagchi:2023cfp, Bagchi:2024rje}. We then analysed the dynamics of Carroll strings in the near horizon region of the  BTZ black hole by performing a $\epsilon$-expansion of the relativistic actions. This procedure yields two distinct Carrollian string theories, depending on the scaling of the relativistic Lagrange multiplier $e$: the magnetic Carroll string with a Lorentzian worldsheet and the electric Carroll string with a Carrollian worldsheet. Equipped with this machinery, we looked at the string dynamics in the near-horizon regime of static and rotating non-extremal BTZ black hole. 

\medskip

In the magnetic case, we solved the expanded Polyakov action upto NLO involving some ansatze and discovered three classes of solution: one in which the string shrinks to a point on ($S^1$), corresponding to null geodesics in the two-dimensional Rindler spacetime; and another two classes involving folded string configurations. One of the folded string configurations shows the complete freezing of string dynamics near the horizon of a static BTZ black hole, whereas the same class of solution in rotating BTZ shows only a trivial dynamics in $\phi$ direction due to frame dragging. For the other class of folded string solution, we found that the string looses its radial width as it approaches the horizon and asymptotes to the null Rindler geodesic (along with the trivial $\phi$ dynamics for rotating BTZ). We also tried to understand the dynamics of these strings from a geometric viewpoint, since they can change size as they are infalling/outgoing w.r.t the horizon. For a robust cross-check, we explicitly established the equivalence of Polyakov approach and phase space approach to magnetic string in near-horizon static BTZ background.
\medskip

In contrast, the electric Carroll string, having an explicit Carroll invariant worldsheet, admits solutions where the string could expand and wrap the transverse circle ($S^1$). Due to the constrained topology of 3D BTZ black hole, the string wrapping mechanism is much more restricted than one expects in a 4D black hole, and this mechanism, described in figure \eqref{fig: electric strings near BTZ Black hole}, gives a multiply wound string around the horizon. The winding of this string could be spatially uniform or non-uniform based on specific configurations we are looking at. 

\subsection*{Future Directions}
The current work opens up several interesting avenues for further exploration. Some of these are described below.   
\begin{itemize}
    \item {\em Extremal Limit:} Our analysis so far has been restricted to the non-extremal BTZ black hole. A natural extension is to study strings in the vicinity of extremal black holes, whose near-horizon geometry is $AdS_2 \times S^1$. This geometry is expected to lead to modifications to the Carrollian structures on the worldsheet. But the string-Carroll formulation of the near-horizon geometry of an extremal black hole requires further analysis. More precisely, the coordinate transformations used to study the near-horizon region of an extremal black hole involve an infinite rescaling of the timelike coordinate. We suspect that due to this rescaling, the near-horizon expansion of the extremal black hole cannot be mapped to the string-Carroll expansion.
    \medskip
    
   The question may be reformulated as whether one should begin with the non-extremal metric and take the extremal and near-horizon limits simultaneously or sequentially. Our analysis indicates that both approaches are possible, but not all coordinate systems are suitable. In particular, we find that Eddington–Finkelstein coordinates are preferable to Boyer–Lindquist coordinates for taking the double limits on non-extremal black holes. Consequently, the near-horizon geometry of an extremal black hole \textit{without the infinite time rescaling} can be mapped to the string-Carroll expansion. But on doing so, we lose the $AdS_2$ factor.    

    \item {\em Connecting to limits of known AdS classical solutions:} To place our results in a broader context, it will be fruitful to relate our results to various limits of well-studied AdS$_3$ classical solutions. The BTZ black hole and its near-horizon limits provide a rich testing ground, but many other geometries, such as warped AdS$_3$, conical defects, and quotient spaces, share structural similarities. By systematically exploring how our near-horizon constructions emerge as specific limits of these known AdS$_3$ solutions, we can uncover deeper geometric and symmetry relations. This approach could also clarify how extremal and flat space limits fit into a broader classification, thereby linking our results to a larger family of spacetimes in 3D gravity.

    \item {\em Generalization to flatspace and FSC:} 
    From here, it is natural to study the flat space limit of the BTZ black hole. The BTZ geometry is locally AdS$_3$ and can be described as an orbifold of AdS$_3$ \cite{PhysRevD.48.1506}. In three-dimensional flat space there are no black holes, but there exist similar solutions called Flat Space Cosmologies (FSCs). These appear when we take the flat space limit of a non-extremal rotating BTZ black hole where the outer event horizon is pushed to infinity. FSCs can also be described as orbifolds of flat space \cite{Cornalba:2002fi, Cornalba:2003kd}. Extending our “String–Carroll” analysis to FSC backgrounds would allow us to explore string dynamics in asymptotically flat spacetimes and possibly uncover links to flat space holography and BMS$_3$ symmetries.

    \item {\em Quantisation and Connecting to worldsheet analysis:} Our analysis of classical string dynamics near the BTZ horizon led to two distinct Carrollian string theories: a magnetic theory with Lorentzian worldsheet and an electric theory with Carrollian worldsheet. An important next step is to connect our spacetime results with a systematic worldsheet analysis. While our current work focuses on the target space description of strings in the near-horizon BTZ geometry, the full string dynamics can be more transparently understood from the worldsheet point of view. Quantising this theory, one could directly probe the string spectrum -- in analogy with \cite{Bagchi:2020fpr}, where three distinct quantum theories were obtained from a single classical null string theory; \cite{Banerjee:2023ekd, Banerjee:2024fbi}, where the spectrum was computed for a constant $B$-field. Beyond the spectrum, such a worldsheet approach provides access to correlation functions and may yield valuable insights into possible dual field-theoretic interpretations. Moreover, the quantum analysis of our target space description may shed light on the quantum spreading of strings across black hole horizons in the tensionless limit — a well-known phenomenon in which an asymptotic observer sees the string diffuse and cover the horizon in finite time \cite{Susskind:1993ki}. This effect has potential implications for understanding information loss during black hole evaporation.

    \item {\em Tensile to Tensionless Strings:} The tensionless limit leads to enhancement of the worldsheet symmetries \cite{Bagchi:2015nca, Bagchi:2024qsb} and collapse of all excitations into massless states. These appear as infinite tower of higher-spin fields \cite{BONELLI2003159}. A natural extension of our analysis is to apply the string-Carroll formalism to study the transition of tensile to tensionless strings in the near-horizon region. While Appendix \ref{appendix:null string} presents explicit family of solutions for tensionless strings in NH region of NE BTZ black hole, the precise connection governing the transition from the tensile solutions remains unknown. Moreover, quantising such tensionless strings may involve subtle, non-standard techniques, as developed in \cite{Bagchi:2024tyq,Dutta:2024gkc}, which could in turn help us understand the worldsheet picture of our Carrollian strings better.

Last but not the least, we should note that Carrollian strings serendipitously appear in an intricate duality web that connects various corners of the Lorentzian/Non-Lorentzian string theory \cite{Blair:2023noj,Gomis:2023eav} (see also \cite{Fontanella:2025tbs, blair2025carrollgeometrymeetssitter, Argando_a_2025} for connections between these corners and Carroll gauge theory), so it is imperative that we understand them and the formalism around them better. 
    We end on the note that String solutions near black holes and the consequences of such dynamics is an absolutely fascinating arena of physics, and it is indeed beautiful that we can access them via the string-Carroll setting.
    We hope to come back to these above-mentioned questions, as well as many more along the same avenue, in future works.
    
\end{itemize}

\subsection*{Acknowledgements}
We are grateful to Arjun Bagchi for valuable discussions and insightful comments on the manuscript, and further for ongoing collaboration on related projects. We also thank Ritankar Chatterjee, Sayan Das, Sachin Grover, Apratim Kaviraj, Nilay Kundu, Shiraz Minwalla, Onkar Parrikar, Srivishnu Rajagopal, Tapobrata Sarkar and Pushkar Soni for helpful discussions. We thank Matthew Headrick for the well-documented \texttt{diffgeo.m} Mathematica package \cite{Headrick:Mathematica}, which we used in some of our computations.
\medskip

Arkachur and SI are supported by IIT Kanpur Institute Assistantship for PhD. ABan is supported in part by an early career research grant ANRF/ECRG/2024/002604/PMS
from ANRF India. He also acknowledges financial support from the Asia Pacific Center for Theoretical Physics (APCTP) via an Associate Fellowship.  AM would like to acknowledge the support of IISER Mohali for the opportunity of an off-campus MSc thesis. Further, he particularly wants to thank BITS Pilani, Pilani campus for kind hospitality during the course of this project, supported by an OPERA grant and a seed grant NFSG/PIL/2023/P3816 from BITS-Pilani. PP acknowledges support from the Infosys Endowment for the Study of the Quantum Structure of Spacetime at TIFR, and earlier from an IIT Kanpur Institute Assistantship for Postdoctoral Fellow. %Most of this work was carried out at IIT Kanpur, whose hospitality is gratefully acknowledged.

\appendixpage
\appendix

\section{Null string solutions} \label{appendix:null string}
In this section, we will give a small glimpse of the consequences of dialling the tension $\hat T$ to zero. This situation would describe null strings in a string-Carroll background, especially in the Electric sector. We start with \eqref{L:general} and set $\hat T = 0$, which yields the null string action \cite{Isberg:1993av}
\begin{equation}\label{L: ILST longitudinal}
     \hat L_{\text{Null}} = \frac{1}{2} \oint  \mathbb{e} \tau_{\mu\nu}\mathbb q^a\partial_a x^\mu \mathbb q^b\partial_b x^\nu \, d\sigma^1.
\end{equation}
We will restrict our analysis to the static BTZ black hole, since the analysis can easily be carried over to the rotating case in the co-rotating coordinates. The null string action near a static BTZ black hole takes the following form
\begin{equation}
        \hat L_{\text{LO}|\text{Null}} = \oint \frac{1}{2\hat{e}} \left[\left(-\frac{x^\rho}{\a}\right)(\xd^t-ux'^t)^2 + \frac{\a}{ x^\rho}(\xd^\rho - ux'^\rho)^2\right] \, d\sigma^1
\end{equation}
To find the equations of motion, we employ the same gauge choices and tilded coordinates as in the case of electric strings of section \eqref{section: static BTZ electric}. In this case, the equations of motion for $u$, $\hat{e}$, $x^t$, $x^\rho$ are a much simplified set: 
\begin{subequations} \label{EOM: Null BTZ New}
    \begin{align}
        0 &= \frac{x^\rho}{\a} \xd^t x'^t - \frac{\a}{x^\rho} \xd^\rho x'^\rho,  \label{EOM:Null BTZ 1}\\
        0 &= \frac{x^\rho}{\a} (\xd^t)^2 - \frac{\a}{x^\rho}(\xd^\rho)^2,  \label{EOM:Null BTZ 2}\\
        0 &= \tilde \partial_0 \left(\frac{x^\rho}{\a}\xd^t\right) \label{EOM:Null BTZ 3} \\
        0 &= \frac{1}{2\a} (\xd^t)^2 + \frac{\a}{2(x^\rho)^2}(\xd^\rho)^2 + \tilde \partial_0 \left(\frac{\a}{x^\rho} \xd^\rho \right).\label{EOM:Null BTZ 4}
    \end{align}
    \end{subequations}

Solving \eqref{EOM: Null BTZ New}, we get the following equations for the LO longitudinal embedding coordinates
\begin{eqnarray}\label{eq:solutions for null strings}
    x^\rho=\pm\frac{\tilde\sigma^0}{\a}f(\tilde\sigma^1)+g(\tilde\sigma^1),~~~x^t=\pm\a\log\left|\frac{\a x^\rho}{f(\tilde\sigma^1)}\right|.
\end{eqnarray}
The functions $f(\tilde\sigma^1)$ and $g(\tilde\sigma^1)$ are integration constants and depend on the initial conditions. However, notice that if they were to be constants, we would obtain the null Rindler geodesic, which we derived for the magnetic Carroll theory at the NLO \eqref{static rindler geodesic}.

\section{Particle Carroll expansion of relativistic strings}\label{app:Particle Carroll expansion of relativistic strings}

In this appendix, we review the particle Carroll expansion of the target spacetimes and explore the equations of motion and the Virasoro constraints of strings on such geometries.

\subsection{Particle-Carroll expansion of target space}
In the particle-Carroll expansion of the target spacetime geometry, the $c$ expansion is carried out only along the timelike coordinate. Therefore, only the timelike coordinate is singled out. Consider a $(d+1)$-dimensional Lorentzian manifold with metric $g_{\mu\nu}$. Similar to the pre-Carrollian variables in the string-Carroll case, the metric and its inverse in the case of the particle-Carroll expansion can be written in terms of the pre-ultra local variables \cite{Hansen:2021fxi} as follows
\begin{equation} \label{appBeq: PUL decom}
    g_{\mu\nu} = -c^2 \mathcal{T}_\mu \mathcal{T}_\nu +  \Psi_{\mu\nu}\,, \,\,\,\,g^{\mu\nu} =  -\frac{1}{c^2}\mathcal{V}^\mu\mathcal{V}^\nu + \Psi^{\mu\nu}.
\end{equation}
The transverse components $\Psi_{\mu\nu}$ and $\Psi^{\mu\nu}$ can be decomposed in terms of the transverse vielbeine in the following way 
\begin{equation}
    \Psi_{\mu\nu} = \delta_{AB}\mathscr{E}^A_\mu\mathscr{E}^B_\nu\,, \,\,\,\,\,\,\,\, \Psi^{\mu\nu} = \delta^{AB}\mathscr{E}^\mu_A\mathscr{E}^{\nu}_B\,,
\end{equation}
where the upper case latin indices now range over $A, B = 1,\ldots, d$. These pre-ultra local variables can be expanded in powers of $c^2$ as
\begin{subequations}
    \begin{align}
        \mathcal{V}^\mu &= \upsilon^\mu + c^2 K^\mu + \mathcal{O}(c^4), \\
        \mathcal{T}_{\mu} &= \uptau_\mu + \mathcal{O}(c^2), \\
        \mathscr{E}^\mu_A &= e^\mu_A + \mathcal{O}(c^2), \\
        \mathscr{E}_\mu^A &= e^A_\mu + c^2 \pi^A_\mu + \mathcal{O}(c^4).
    \end{align}
\end{subequations}

These expansions, when substituted in equation \eqref{appBeq: PUL decom}, give the particle Carroll expansion of the target spacetime metric as follows 
\begin{align} \label{appBeq:PCE}
        g_{\mu\nu} &= \Omega_{\mu\nu} - c^2 \uptau_\mu\uptau_\nu + c^2 \Theta_{\mu\nu} + \mathcal{O}(c^4) \nn \\
        g^{\mu\nu} &= -\frac{1}{c^2}\upsilon^\mu\upsilon^\nu + \Tilde{\Omega}^{\mu\nu} + \mathcal{O}(c^2), 
\end{align}
where,
\begin{equation*}
    \Theta_{\mu\nu} = 2\delta_{AB}e^A_{(\mu}\pi^B_{\nu)}\,, \,\,\,\,\,\,\,\, \tilde{\Omega}^{\mu\nu} = \Omega^{\mu\nu} - 2\upsilon^{(\mu}K^{\nu)}\,.
\end{equation*}

\subsection{Magnetic strings in the particle Carroll geometry}
Analogous to section \ref{Polyakov action in the string Carroll metric}, here, we explore the particle-Carroll expansion of the Polyakov action and determine the classical dynamics of strings in this background.
\subsubsection{Expansion of the Polyakov action}
We consider the following ansatz to expand the dynamical variables of the Polyakov action \eqref{Polyakov} in a power series in $c^2$ 
\begin{subequations}
    \begin{align}
        X^\m &= x^\m + c^2 y^\m + \O(c^4)\,, \\
        \gamma_{ab} &= \gamma_{(0) ab} + c^2 \gamma_{(2)ab} + \O(c^4) \,, \\
        \gamma^{ab} &= \gamma_{(0)}^{ab} + c^2 \gamma_{(2)}^{ab} + \O(c^4) \,.
    \end{align}
\end{subequations}

Clearly, $\gamma_{(0)}$ is Lorentzian and $\gamma_{(0)ab}\gamma_{(0)}^{bc} = \delta^c_a$. The induced metric on the worldsheet $g_{ab}(X) = g_{\m\n}(X)\partial_a X^\m \partial_b X^\n$ expands as
\begin{equation}
    g_{ab}(X) = \Omega_{ab}(x) + c^2 \hat\Theta_{ab}(x, y) + \O(c^4) \,,
\end{equation}

where,
\begin{subequations}
    \begin{align}
        \hat\Theta_{ab}(x,y) &= \uptau_{ab}(x) + \Theta_{ab}(x) + 2 \Omega_{\m\n}(x) \partial_{(a}x^\m \partial_{b)}y^\n \nn\\
        &\qquad \qquad \qquad \quad + \partial_a x^\m \partial_b x^\n y^\lambda \partial_\lambda \Omega_{\m\n}(x) \,, \\
        \uptau_{ab}(x) & = \partial_a x^\m \partial_b x^\n \uptau_\m(x) \uptau_\n (x) \,,
        \\\Theta_{ab}(x)&= \partial_a x^\m \partial_b x^\n \Theta_{\m\n}(x)\,,\\
        \Omega_{ab}(x) &= \partial_a x^\m \partial_b x^\n \Omega_{\m\n}(x) \,.
    \end{align}    
\end{subequations}

The expanded Polyakov action takes the form
\begin{equation}
    S_P = S_{P,LO} + S_{P,NLO} \,,
\end{equation}
where
\begin{subequations}\label{eq: polyakov action particle carroll}
    \begin{align}
        S_{P,LO} &= -\frac{T}{2} \int \Omega_{ab} \gamma_{(0)}^{ab} \lb x \rb \sqrt{-\gamma_{(0)} } \, d^2\xi \,, \label{eq: LO polyakov action particle carroll} \\
        S_{P,NLO} &= -\frac{T}{2} \int \left[ \gamma_{(0)}^{ab}\hat\Theta_{ab} \lb x, y \rb - \frac{1}{2} G^{abcd}_{(0)}\Omega_{ab} \lb x \rb \gamma_{(2)cd} \right] \sqrt{-\gamma_{(0)} } \, d^2 \xi \, , \label{eq: NLO polyakov action particle carroll}
    \end{align}
\end{subequations}
where $ G^{abcd}_{(0)} = \gamma_{(0)}^{ac} \gamma_{(0)}^{bd} + \gamma_{(0)}^{ad} \gamma_{(0)}^{bc} - \gamma_{(0)}^{ab} \gamma_{(0)}^{cd} $ is the Wheeler-DeWitt metric. For meaningful ways to solve the theories, we will first look at worldsheet symmetries, which will help us simplify the theory via gauge fixing conditions.

\subsubsection*{Worldsheet symmetries of particle Carroll Polyakov action}

Under infinitesimal diffeomorphisms $\xi^a \to \xi^a + \zeta^a$ and local Weyl rescaling $\chi$, the worldsheet metric transforms as
\begin{equation}\label{eq: particle carroll worldsheet infitesimal}
	\delta \gamma_{ab} = \mathcal{L}_\zeta \gamma_{ab} + 2 \chi\gamma_{ab}\,,
\end{equation}
We now postulate that $\zeta^a$ and $\chi$ has an analytical expansion in powers of $c^2$ as follows
\begin{subequations}
    \begin{align}
        \zeta^a &= \zeta_{(0)}^a + c^2 \zeta_{(2)}^a + \O(c^4) \,, \\
        \chi &= \chi_{(0)} + c^2\chi_{(2)} + \O(c^4) \,.
    \end{align}
\end{subequations}

Thus we get $\delta \gamma_{ab} = \delta \gamma_{(0)ab} + c^2 \delta \gamma_{(2)ab} + \O(c^4)$, where
\begin{subequations}
    \begin{align}\label{eq: LO infinitesimal expansion}
        \delta \gamma_{(0)ab} & = \mathcal{L}_{\zeta_{(0)}} \gamma_{(0)ab} + 2\chi_{(0)}\gamma_{(0)ab} \,, \\
		\label{eq: NLO infinitesimal expansion}
		\delta \gamma_{(2)ab} & = \mathcal{L}_{\zeta_{(0)}} \gamma_{(2)ab} + \mathcal{L}_{\zeta_{(2)}} \gamma_{(0)ab} + 2\chi_{(0)}\gamma_{(2)ab} + 2\chi_{(2)}\gamma_{(0)ab} \,.
    \end{align}
\end{subequations}

From the similarity between the structures of \eqref{eq: LO infinitesimal expansion} and \eqref{eq: particle carroll worldsheet infitesimal}, and remembering that $\gamma_{ab}$ can be gauge fixed to a Minkowski metric, locally, one can fix the gauge $\gamma_{(0)ab} = \eta_{ab}$. Moreover, for this choice of gauge, one can choose parameters $\zeta_{(2)}$ and $\chi_{(2)}$ that set $\gamma_{(2)} = 0$. These two gauge-fixing conditions help to simplify the constraint equations and equations of motion. Thus, we will employ this gauge throughout the remainder of the analysis.

\subsubsection*{LO theory}

Varying $\gamma^{ab}_{(0)}$ in \eqref{eq: LO polyakov action particle carroll} gives us the LO Virasoro constraints 
\begin{equation}\label{eq: LO virasoro constraint particle carroll}
  T_{(0)ab} \lb x \rb := \Omega_{ab} \lb x \rb - \frac 1 2 \gamma_{(0)}^{cd}\Omega_{cd} \lb x \rb \gamma_{(0)ab} = 0 \,,
\end{equation}

and the LO equations of motion for $x^\m$  are
\begin{equation}\label{eq: LO eom particle carroll}
    \frac{1}{2}\sqrt{ - \gamma_{(0)} } \gamma_{(0)}^{ab} \partial_a x^\nu \partial_b x^\lambda \partial_\mu \Omega_{\nu\lambda} = \partial_a \left(  \sqrt{ - \gamma_{(0)} } \gamma_{(0)}^{ab} \partial_b x^\nu \Omega_{\mu\nu} \right)\,.
\end{equation}

Using the gauge fixing condition, that is, $\gamma^{ab}_{(0)}=\eta^{ab}$, the Virasoro constraints \eqref{eq: LO virasoro constraint particle carroll} take the form
\begin{subequations} \label{eq: gf virasoro constraint particle carroll}
    \begin{align}
        \partial_0 x^\m \partial_0 x^\n \Omega_{\m\n} \lb x \rb + \partial_1 x^\m \partial_1 x^\n \Omega_{\m\n} \lb x \rb &= 0 \,, \label{eq: LO gf virasoro constraint particle carroll} \\
        \partial_0 x^\m \partial_1 x^\n \Omega_{\m\n} \lb x \rb &= 0 \,. \label{eq: NLO gf virasoro constraint particle carroll}
    \end{align}
\end{subequations}

Since $\Omega_{\m\n}$ is positive semidefinite, equations \eqref{eq: gf virasoro constraint particle carroll} imply
\begin{equation}\label{eq: LO particle carroll sol}
    \Omega_{\m\n} \lb x \rb \partial_a x^\n  = 0 \,.
\end{equation}

It is easy to verify that \eqref{eq: LO particle carroll sol} trivially solves the LO equation of motion \eqref{eq: LO eom particle carroll}. Now we will move on to the NLO Polyakov theory.

\subsubsection*{NLO theory}

Varying the NLO Polyakov action \eqref{eq: NLO polyakov action particle carroll} with respect to $\gamma_{(2)ab}$ gives back the LO Virasoro constraint \eqref{eq: LO virasoro constraint particle carroll}. Whereas, varying $S_{P,NLO}$ with respect to $\gamma^{ab}_{(0)}$ gives the NLO Virasoro constraint
\begin{equation}
    T_{(2)ab} := \hat\Theta_{ab}\lb x, y \rb - \frac{1}{2}\gamma_{(0)}^{cd}\hat\Theta_{cd}\lb x, y \rb \gamma_{(0)ab} + \left(\text{terms with } \gamma_{(2)} \right) = 0 \,.
\end{equation}

The variation of the NLO Polyakov action \eqref{eq: NLO polyakov action particle carroll} with respect to $y^\m$ gives the LO equations of motion for $x^\m$ \eqref{eq: LO eom particle carroll}. The equations of motion for $x^\m$ are lengthy and do not offer much physical insight. So we skip them here.

The gauge-fixed NLO Virasoro constraints, along with the LO solution \eqref{eq: LO particle carroll sol}, imply
\begin{subequations}\label{eq: NLO gf virasoro part 1}
    \begin{align}
        \delta^{ab} \left( \uptau_{ab} \lb x \rb + \Theta_{ab} \lb x \rb  \right) &= 0\,, \\
        \uptau_{01} \lb x \rb + \Theta_{01} \lb x \rb & = 0 \,.
    \end{align}
\end{subequations}

This concludes our NLO analysis of the particle Carroll expansion of the Polyakov string action. 
%For detailed solutions on how strings behave near particle Carroll geometries, one needs to have an explicit example. Those solutions will have some specific geometry-dependent features along with these universal features.

\subsubsection{The phase space formalism} 
%start editing from here
To obtain the magnetic sector from the relativistic phase space Lagrangian \eqref{eq:relativistic phase space action}, we first rescale $e \rightarrow \tilde{e}:=c^2e$ and $T \rightarrow \tilde{T}:= cT$. Now, expanding the embedding fields, the momenta and the Lagrange multipliers, as done in section 
\ref{Phase space analysis for magnetic string}, we get a $c^2$-expansion of the Lagrangian 

\begin{equation}
    L = \tilde{L}_{LO} + c^2 \tilde{L}_{NLO} + + c^4 \tilde{L}_{NNLO}+\mathcal{O}(c^6),
\end{equation}
where,
\begin{subequations}
    \begin{align}
    \tilde{L}_{LO} &= \frac{1}{2}\oint d\sigma \, \tilde{e}_{(0)} \upsilon^{\mu\nu}P_{(0)\mu}P_{(0)\nu} \label{appBeq: LO magnetic particle Carroll}\\
    \tilde{L}_{NLO} &= \oint d\sigma  \Big[\dot{x}^\mu P_{(0)\mu}+ \frac{1}{2}\tilde{e}_{(2)}\upsilon^{\mu\nu}P_{(0)\mu}P_{(0)\nu} - u_{(0)}x'^{\mu}P_{(0)\mu} \nn \\
    &~~~~~~~~-\frac{1}{2}\tilde{e}_{(0)}\left(\tilde{\Omega}^{\mu\nu}(x,y)P_{(0)\mu}P_{(0)\nu} + \tilde{T}^2\Omega_{\mu\nu} x'^{\mu}x'^{\nu} - 2\upsilon^{\mu\nu}P_{(2)\mu}P_{(0)\nu}\right)\Big]\,. \label{appBeq: NLO magnetic particle Carroll}\\
    \tilde{L}_{\mathrm{NNLO}}&=\oint d \sigma^1\left[\dot{y}^\mu P_{(0) \mu}+\dot{x}^\mu P_{(2) \mu}-\frac{1}{2} \tilde{e}_{(4)} \upsilon^{\mu} \upsilon^{\nu} P_{(0) \mu} P_{(0) \nu}-\frac{1}{2} \tilde{e}_{(0)}\left\{\Psi^{\mu \nu}(x, y, z) P_{(0) \mu} P_{(0) \nu}\right.\right. \nn \\
& +2 \tilde{\Omega}^{\mu \nu}(x, y) P_{(0) \mu} P_{(2) \nu}+\upsilon^{\mu} \upsilon^{\nu} P_{(2) \mu} P_{(2) \nu}+2 \upsilon^{\mu} \upsilon^{\nu} P_{(4) \mu} P_{(0) \nu}\nn\\&+\tilde{T}^2\left(\uptau_{\mu} \uptau_{\nu}+\Theta_{\mu \nu}(x, y)\right) x^{\prime \mu} x^{\prime \nu}
 \left.+2 \tilde{T}^2 \Omega_{\mu \nu} x^{\prime \mu} y^{\prime \nu}\right\}\nn\\&-\frac{1}{2} \tilde{e}_{(2)}\left(\tilde{T}^2 \Omega_{\mu \nu} x^{\prime \mu} x^{\prime \nu}+\tilde{\Omega}^{\mu \nu}(x, y) P_{(0) \mu} P_{(0) \nu}+2 \upsilon^{\mu} \upsilon^{\nu} P_{(0) \mu} P_{(2) \nu}\right) \nn \\
& \left.-u_{(2)} x^{\prime \mu} P_{(0) \mu}-u_{(0)}\left(y^{\prime \mu} P_{(0) \mu}+x^{\prime \mu} P_{(2) \mu}\right)\right] \label{appBeq: NNLO magnetic particle Carroll}.
\end{align}
\end{subequations}

Therefore, the form of the LO and NLO magnetic string Lagrangians does not change when the particle Carroll expansion is used instead of the string Carroll expansion.

\subsection*{LO theory}

From the LO Lagrangian \ref{appBeq: LO magnetic particle Carroll} one obtains the equations of motion associated with $P_{(0)\mu}$ and $\tilde{e}_{(0)}$ as 
\begin{subequations}
    \begin{align}
        \tilde{e}_{(0)}\upsilon^{\mu\nu}P_{(0)\nu} &= 0 \label{appBeq: magnetic LO Eom1}\\
        \upsilon^{\mu\nu}P_{(0)\mu}P_{(0)\nu} &= 0 \label{appBeq: magnetic LO Eom2}
    \end{align}
\end{subequations}

Assuming that $\tilde{e}_{(0)} \neq 0$, the equation \ref{appBeq: magnetic LO Eom1} gives 
\[\upsilon^{\mu\nu}P_{(0)\nu} = 0,\]

which trivially satisfies equation \ref{appBeq: magnetic LO Eom2} as well. Therefore, the LO theory is trivial.

\subsection*{NLO theory}
The NLO Lagrangian \ref{appBeq: NLO magnetic particle Carroll} gives the same equation of motion for $P_{(2)\mu}$ as that for $P_{{(0)}\mu}$ in the LO theory. Substituting it into the Lagrangian, we get
\begin{equation}
    \tilde{L}_{NLO} = \oint d\sigma \Bigg[\dot{x}^\mu P_{(0)\mu}-\frac{1}{2}\tilde{e}_{(0)}\left(\Omega^{\mu\nu}(x)P_{(0)\mu}P_{(0)\nu} + \tilde{T}^2\Omega_{\mu\nu} x'^{\mu}x'^{\nu} \right) - u_{(0)}x'^{\mu}P_{(0)\mu}\Bigg]
\end{equation}

Varying the above action with respect to $P_{(0)\mu}$ we get, 
\begin{equation}\label{eq:EOM P0 particle Carroll}
    \Omega^{\mu\nu}P_{(0)\nu} = \frac{1}{\tilde{e}_{(0)}} \Omega^\mu_\nu\left(\dot{x}^\nu - u_{(0)}x'^{\nu}\right).
\end{equation}

Substituting \eqref{eq:EOM P0 particle Carroll} into $\tilde{L}_{NLO}$, we get, 
 \begin{equation}
     \tilde{L}_{NLO} = \oint d\sigma \frac{1}{\tilde{e}_{(0)}}\left[\Omega_{\mu\nu}\dot{x}^\mu \dot{x}^{\nu} + (u_{(0)}^2 - \tilde{e}_{(0)}^2 \tilde{T}^2)\Omega_{\mu\nu}x'^{\mu}x'^{\nu} - 2u_{(0)}\Omega_{\mu\nu}\dot{x}^\mu x'^{\nu}\right]
 \end{equation}

 The equations of motion for $\tilde{e}_{(0)}$, $u_{(0)}$ and $x^\mu$ are 
\begin{subequations}
    \begin{align}
    \Omega_{\mu\nu}x'^{\mu}\left(\dot{x}^\nu -  u_{(0)}x'^{\nu}\right) &= 0 \\
    \Omega_{\mu\nu}\left[\dot{x}^\mu \dot{x}^{\nu} + (u_{(0)}^2 + \tilde{e}_{(0)}^2\tilde{T}^2) x'^{\mu}x'^{\nu} - 2u_{(0)}\dot{x}^\mu x'^{\nu}\right] &= 0 \\
    \left[\dot{x}^\mu \dot{x}^{\nu} + (u_{(0)}^2 - \tilde{e}_{(0)}^2 \tilde{T}^2)x'^{\mu}x'^{\nu} - 2u_{(0)}\dot{x}^\mu x'^{\nu}\right]\partial_{\rho}\Omega_{\mu\nu} -\nn \\ \partial_{\tau}\left(2\Omega_{\rho\nu}\dot{x}^{\nu} - 2u_{(0)}\Omega_{\rho\nu}x'^{\nu}\right) - \partial_{\sigma}\left(2(u_{(0)}^2 - \tilde{e}_{(0)}^2 \tilde{T}^2)\Omega_{\rho\nu}x'^{\nu} - 2u_{(0)}\Omega_{\rho\nu}\dot{x}^{\nu}\right) &= 0
    \end{align}
\end{subequations}

Fixing the gauge to $\tilde{e}_{(0)}\tilde{T} = 1$ and $u_{(0)} = 0$ (locally) simplifies the equations of motion as

\begin{subequations}
    \begin{align}
        \Omega_{\mu\nu}\dot{x}^\mu x'^{\nu} &= 0 \\
        \Omega_{\mu\nu}\left(\dot{x}^\mu \dot{x}^{\nu} + x'^{\mu} x'^{\nu}\right) &= 0 \\
         \left[\dot{x}^\mu \dot{x}^{\nu} + x'^{\mu}x'^{\nu}\right]\partial_{\rho}\Omega_{\mu\nu} - 2\partial_{\tau}\left(\Omega_{\rho\nu}\dot{x}^{\nu} \right) + 2\partial_{\sigma}\left(\Omega_{\rho\nu}x'^{\nu}\right) &= 0
    \end{align}
\end{subequations}

The above equations of motion describe the dynamics of the string in the particle Carroll expansion of the target spacetime metric. 

\subsubsection*{NNLO theory}
 Integrating out all the momentum fields from the NNLO Lagrangian \ref{appBeq: NNLO magnetic particle Carroll} we get
\begin{equation}
    \begin{aligned}
\tilde{L}_{\mathrm{NNLO}}= & \oint d \sigma^1 \frac{1}{2 \tilde{e}_{(0)}}\left[\left(\dot{x}^\mu \dot{x}^\nu-2 u_{(0)} \dot{x}^\mu x^{\prime \nu}+\left(u_{(0)}^2-\tilde{e}_{(0)}^2 \tilde{T}^2\right) x^{\prime \mu} x^{\prime \nu}\right)\left(\uptau_{\mu} \uptau_{\nu}+\Theta_{\mu \nu}(x, y)\right)\right. \\
& +2 \Omega_{\mu \nu}\left(\dot{x}^\mu \dot{y}^\nu-2 u_{(0)} \dot{x}^\mu y^{\prime \nu}+\left(u_{(0)}^2-\tilde{e}_{(0)}^2 \tilde{T}^2\right) x^{\prime \mu} y^{\prime \nu}\right) \\
& +\Omega_{\mu \nu} \frac{\tilde{e}_{(2)}}{\tilde{e}_{(0)}}\left(-\dot{x}^\mu \dot{x}^\nu+2 u_{(0)} \dot{x}^\mu x^{\prime \nu}+\left(-u_{(0)}^2-\tilde{e}_{(0)}^2 \tilde{T}^2\right) x^{\prime \mu} x^{\prime \nu}\right) \\
& \left.+2 \Omega_{\mu \nu} u_{(2)}\left(-\dot{x}^\mu \dot{x}^\nu+u_{(0)} x^{\prime \mu} x^{\prime \nu}\right)\right] .
\end{aligned}
\end{equation}

The equations of motion corresponding to $\tilde{e}_{(0)}, u_{(0)}, x^\mu, \tilde{e}_{(2)}, u_{(2)}$ and $y^{\mu}$ are rather lengthy and do not offer additional insight, so we refrain from presenting them explicitly here.

\subsection{Electric strings in particle Carroll geometry}
The electric sector is now obtained by rescaling $e \rightarrow\hat{e} := e$ and $T \rightarrow \hat{T} := c^2 T$ in \eqref{eq:relativistic phase space action}. Expanding the embedding fields, momenta and Lagrange multipliers in powers of $c^2$, we get the electric $c^2$-expansion of the relativistic Lagrangian as 
\begin{equation}
    L = \hat{L}_{LO} + \mathcal{O}(c^2),  
\end{equation}
where,
\begin{equation}
    \hat{L}_{LO} = \oint d \sigma^1 \left[\dot{x}^\mu P_{(0)\mu} + \frac{\hat{e}_{(0)}}{2}\left(\upsilon^\mu \upsilon^\nu P_{(0)\mu}P_{(0)\nu} - \hat{T}^2 \Omega_{\mu\nu}x'^\mu x'^\nu \right) - u_{(0)}x'^\mu P_{(0)\mu}\right]\,.
\end{equation}
Integrating out the longitudinal part of the momentum, we get
\begin{align} \label{appBeq: LO electric string}
    \hat{L}_{LO} = \oint d \sigma^1 \Bigg[\left(\dot{x}^\mu - u_{(0)}x'^\mu\right)e^{\bar{A}}_\mu P_{\bar{A}} - \frac{1}{2\hat{e}_{(0)}} \uptau_\mu \uptau_\nu \big(\dot{x}^\mu - u_{(0)}&x'^\mu\big)\left(\dot{x}^\nu - u_{(0)}x'^\nu\right) \nn \\ 
    &- \frac{\hat{T}^2}{2}\hat{e}_{(0)}\Omega_{\mu\nu}x'^\mu x'^\nu\Bigg]\,,
\end{align}
where, $P_{\bar{A}}$ denotes the transverse momenta. The equations of motion for $P_{\bar{A}}$, $u_{(0)}$, $e_{(0)}$ and $x^\mu$ are as follows

\begin{subequations}
\begin{align}
\left(\dot{x}^\mu - u_{(0)}x'^\mu\right)e^{\bar{A}}_\mu &= 0 \\
-x'^{\mu}e_{\mu}^{\bar{A}}P_{\bar{A}}
 + \frac{1}{\hat{e}_{(0)}} \uptau_\mu \uptau_\nu x'^{\mu} 
   \left(\dot{x}^\nu - u_{(0)}x'^{\nu}\right) &= 0 \\
\frac{1}{2\hat{e}_{(0)}^2}\uptau_\mu \uptau_\nu
 \left(\dot{x}^\mu - u_{(0)}x'^\mu\right)
 \left(\dot{x}^\nu - u_{(0)}x'^\nu\right)
 - \frac{\hat{T}^2}{2}\Omega_{\mu\nu}x'^\mu x'^\nu &= 0 \\
\left(\dot{x}^\mu - u_{(0)}x'^\mu\right) P_{\bar{A}} 
   \partial_{\rho} e^{\bar{A}}_{\mu} 
- \frac{1}{2\hat{e}_{(0)}} 
   \left(\dot{x}^\mu - u_{(0)}x'^\mu\right)
   \left(\dot{x}^\nu - u_{(0)}x'^\nu\right) 
   \partial_{\rho}(\uptau_\mu \uptau_\nu) \nn \\
- \frac{\hat{T}^2}{2}\hat{e}_{(0)} x'^{\mu} x'^{\nu} 
   \partial_{\rho}(\Omega_{\mu\nu}) + \partial_{\sigma}\Bigg[
      u_{(0)} e^{\bar{A}}_\rho P_{\bar{A}}
      - \frac{u_{(0)}}{\hat{e}_{(0)}} 
        \uptau_{\rho}\uptau_{\mu}
        \left(\dot{x}^\mu - u_{(0)}x'^{\mu}\right) \nn \\
\qquad\quad
      - \frac{\hat{T}^2}{2} \hat{e}_{(0)} \Omega_{\rho\mu} x'^\mu
   \Bigg]
- \partial_{\tau}\Bigg[
      e^{\bar{A}}_\rho P_{\bar{A}}
      - \frac{1}{\hat{e}_{(0)}} \uptau_{\rho}\uptau_{\mu} 
        \left(\dot{x}^\mu - u_{(0)}x'^\mu\right)
   \Bigg] &= 0
\end{align}
\end{subequations}
Fixing the gauge $\hat{e}_{(0)} = 1$, $u_{(0)} = u_{(0)}(\tau)$ and performing the reparametrisation as in \eqref{eq: electric string local gauge} simplifies the equations of motion as
\begin{subequations}
\begin{align}
\dot{x}^\mu e^{\bar{A}}_\mu &= 0\,, \\
\uptau_{\mu}\uptau_{\nu}x'^{\mu}\dot{x}^\nu-x'^{\mu}e_{\mu}^{\bar{A}}P_{\bar{A}} &= 0\,, \\
\uptau_{\mu}\uptau_{\nu}\dot{x}^{\mu}\dot{x}^{\nu}-\hat{T}^2 \Omega_{\mu\nu}x'^\mu x'^\nu &= 0\,, \\
\dot{x}^{\mu}P_{\bar{A}}\partial_{\rho}e^{\bar{A}}_{\mu}-\frac{1}{2}\dot{x}^{\mu}\dot{x}^{\nu}\partial_{\rho}(\uptau_{\mu}\uptau_{\nu})-\frac{\hat{T}^2}{2}x'^{\mu}x'^{\nu}\partial_{\rho}(\Omega_{\mu\nu})\nn\\
-\partial_{\tilde{\tau}}\!\left[e^{\bar{A}}_\rho P_{\bar{A}}-\uptau_{\rho}\uptau_{\mu}\dot{x}^\mu \right]+ \frac{\hat{T}^2}{2}\partial_{\tilde{\sigma}}\left(\Omega_{\rho\mu}x'^\mu\right)&=0\,.
\end{align}
\end{subequations}
\medskip

In summary, the particle–Carroll expansion of the target spacetime offers a natural starting point for extending similar techniques to explore string behaviour within wider contexts than the near-horizon dynamics. This approach provides a systematic framework for investigating string dynamics in various settings, such as near null infinity or on constant radial slices of black hole backgrounds, where the perturbative expansion of the target space naturally manifests a particle–Carroll geometric structure.

\newpage
\bibliographystyle{JHEP.bst}
\bibliography{bibliography}
\end{document}